\documentclass[10pt,epsfig]{article} 
\pdfoutput=1
\usepackage{pstricks}
\usepackage{enumerate}
\usepackage{float}
\usepackage{subfig}
\usepackage{color}
\usepackage{slashed}
\usepackage{amssymb, amsmath,mathrsfs}
\usepackage{graphicx}
\usepackage{multirow}
\usepackage{cite}
\usepackage{chngcntr}
\usepackage[colorlinks=true,
            linkcolor=red,
            urlcolor=blue,
            citecolor=blue]{hyperref}
\usepackage{dcolumn}
\usepackage{bm}
\usepackage{color}
\long\def\rpl#1!!#2!!{\textcolor{red}{#1} \textcolor{blue}{#2}}
\def\baselinestretch{1.3}
\newcommand{\ba}{\begin{array}}
\newcommand{\ea}{\end{array}}
\newcommand{\bd}{\begin{displaymath}}
\newcommand{\ed}{\end{displaymath}}
\newcommand{\besub}{\begin{subequations}}
\newcommand{\eesub}{\end{subequations}}
\newcommand{\be}{\begin{equation}}
\newcommand{\ee}{\end{equation}}
\newcommand{\bea}{\begin{eqnarray}}
\newcommand{\eea}{\end{eqnarray}}
\newcommand{\no}{\nonumber\\}

\def\a{\alpha}

\def\b{\beta}

\def\l{\lambda}
\def\m{\mu}

\def\L{\Lambda}

\def\q2 {q^2}

\def\bt{\begin{table}}
\def\et{\end{table}}

\catcode`@=11 % This allows us to modify PLAIN macros.
\def \gsim{\mathrel{\mathpalette\@versim>}}
\def \lsim{\mathrel{\mathpalette\@versim<}}
\def \@versim#1#2{\lower0.4ex\vbox{\baselineskip\z@skip\lineskip\z@skip
     \lineskiplimit\z@\ialign{$\m@th#1\hfil##\hfil$
     \crcr#2\crcr\sim\crcr}}}
\catcode`@=12 % at signs are no longer letters

\allowdisplaybreaks
\begin{document}

%THE TEXT STARTS HERE
\begin{flushright}
{HRI-RECAPP-2015-017}
\end{flushright}

\begin{center}

{\large \textbf {High-scale validity of a model with Three-Higgs-doublets}}\\[15mm]

Nabarun Chakrabarty$^{\dagger}$\footnote{nabarunc@hri.res.in}  \\
$^{\dagger}${\em Regional Centre for Accelerator-based Particle Physics \\
     Harish-Chandra Research Institute\\
 Chhatnag Road, Jhunsi, Allahabad - 211 019, India}\\[5mm] 

\end{center}

\begin{abstract} 

% We consider the conditions for the validity of a two-Higgs doublet model
%at high energy scales, together with all other low- and high-energy
%constraints. The constraints on the parameter space at low energy,
%including the measured value of the  Higgs mass and the signal strengths
%in channels are juxtaposed with the conditions of vacuum stability,
%perturbativity and unitarity at various scales. We find that a scenario with
%an exact $\mathbb{Z}_2$ symmetry in the potential cannot be valid beyond about
%10 TeV without the intervention of additional physics. On the other hand,
%when the $\mathbb{Z}_2$ symmetry is broken, the theory can be valid even up to the
%Planck scale without any new physics coming in. The interesting feature we
%point out is that such high-scale validity is irrespective of the uncertainty in the top
%quark mass as well as $\a_s(M_Z)$, in contrast with the standard model with a single Higgs doublet. It is also shown that the presence of a CP-violating phase is
%allowed when the $\mathbb{Z}_2$ symmetry is relaxed. The allowed regions in the
%parameter space are presented for each case. The results are illustrated in the context of a Type-II scenario.

We consider a three-Higgs doublet scenario in this paper, invariant under the discrete group $S_3$,
and probe its high-scale validity by allowing the model parameters to evolve under renormalisation
group. We choose two particular alignments of vacuum expectation values (vev) for our study, out of 
a set of several such possible ones. All three doublets receive non-zero vacuum expectation values
in the first case, and in the second case, two of the doublets remain without vev. The constraints on the parameter space at low energy, including the measured value of the Higgs mass and the signal strengths,
oblique corrections and also measurements of relic density and direct detection rates
are juxtaposed with the conditions of vacuum stability, perturbativity and unitarity at various scales.
We find that the scenario with three non-zero vevs is not valid beyond $10^7$ GeV, assuming no  
additional physics participates at the intermediate scales. On the contrary, the scenario with only 
one non-zero vev turns out to be a successful model for cold dark matter phenomenology, which also
turns out to be valid up to the Planck scale at the same time. Stringent restrictions are obtained on
the model parameter space in each case. Thus, the $S_3$ symmetric
scalar sector emerges as an ultraviolet (UV) complete theory.

\end{abstract}

\newpage
\setcounter{footnote}{0}

\def\baselinestretch{1.5}
\counterwithin{equation}{section}
%==========================================================================
%==========================================================================
\section{Introduction}\label{Intro}

The discovery of a scalar boson around 125 GeV\cite{Aad:2012tfa,Chatrchyan:2012ufa} 
has been the most important finding at the LHC so far. It has gradually unfurled that the newly
discovered boson has properties largely consistent with the Standard Model (SM) Higgs\cite{Freitas:2012kw,Djouadi:2013yb,Djouadi:2013qya}. There is
however one pressing issue that a SM Higgs around 125 GeV leads to an instability in the electroweak
vaccum around $10^{8-9}$ GeV if the top quark mass $(M_t)$ and the strong coupling 
constant $(\alpha_s)$ are on the upper edges of their respective uncertainty bands.
A recent next-to-next-to-leading order (NNLO) study \cite{Degrassi:2012ry,Buttazzo:2013uya}
reports that absolute stability up to the Planck scale requires
\begin{eqnarray}
M_h [\rm GeV] > 129.4 + 1.4(\frac{M_t[\rm GeV]-173.1}{0.7})-0.5(\frac{\alpha_s(M_Z)-0.1184}{0.0007})\pm1.0_{th} 
\end{eqnarray}
As a possible remedy to alleviate the vacuum stability problem, the SM Higgs can be made to couple 
to additional bosonic degrees of freedom. In such a case, the extra scalar loops contributing to the 
running of the Higgs coupling can generate the required positive contribution to prevent the Higgs 
self coupling from turning negative. Thus ensuring a stable Elecroweak vacuum (EW) up to the GUT and Planck scales, forms our main motivation to add new scalars to the theory. 
Apart from this, various cosmological and astrophysical evidences of Dark Matter (DM)
also necessiate physics beyond the SM. Attempts have been made to explore the yet 
unknown particle nature of DM and the most successful proposal is that DM is constituted of 
Weakly Interacting Massive Particles (WIMPs)\cite{Bertone:2004pz}.

It is however not possible to predict the actual number of scalar doublets present in nature
from fundamental principles. The discovered scalar resonance around 125 GeV could very
well arise from a multiple scalar doublet scenario with the additional parameters arranged 
suitably to give to it, SM-like couplings to fermions and gauge bosons. Of course, detection
of the extra scalars at the colliders is the only direct way to pin down on the exact 
scalar structure present. Nonetheless these extra scalars could be fingerprinted through 
their role in stabilizing and unitarizing of the scalar potential. Moreover, the scalars
originating from the extra doublets could possible be successful canditates for DM. While
studies combining together vacuum stability and DM phenomenology have been done in the past
in context of two-Higgs doublet models (2HDM)\cite{Goudelis:2013uca}, these could be generalized to a higher number 
of doublets as well. There has been a rising interest in three-Higgs doublet models (3HDM) in 
the recent past\cite{Aranda:2012bv,Moretti:2015tva,Ivanov:2012ry,Maniatis:2015kma,Moretti:2015cwa,Ivanov:2012fp,Pramanick:2015qga,
Keus:2015xya,Keus:2014jha,Keus:2013hya,Varzielas:2015joa}. The chief phenomenological motivation of which is that the existence of
three scalar doublets, replicating the three fermion families, sheds light on the flavor problem. 
3HDMs have rather wide scalar spectrum.  In fact, invariance under SU(2) $\times$ U(1) tells us 
that there are four neutral scalars, and a pair of charged scalars obtainable from a generic 3HDM.
It is reminded that 3HDMs come in various types, depending upon the global symmetry present. One
of them is the 3HDM endowed with a global $S_3$ symmetry\cite{Bhattacharyya:2012ze,Bhattacharyya:2010hp,Barradas-Guevara:2014yoa,Kubo:2004ps,Koide:2005ep,Machado:2012ed}. This $S_3$ symmetry is already important
from the perspective of flavor, it reproduces the lepton masses and mixings accurately~\cite{Harrison:2003aw,Kubo:2003iw,Teshima:2005bk,Koide:2006vs,Chen:2004rr,
Mondragon:2007af}. The scalar 
sector is also interesting since there is an economy
of parameters compared to a more generic 3HDM. In fact the eight dimensionless parameters can be fully traded off in favour of the seven masses and one mixing angle. The $S_3$-symmetric scalar sector
has spurred some investigation in the past, and some standalone studies related to DM phenomenology
have also occurred\cite{Fortes:2014dca}. However, the present study is mainly directed towards
analysing the Higgs sector, and, it
includes the following features which have not been highlighted before.

\begin{itemize}

        \item We derive the renormalisation group equations at one-loop for the
dimensionless parameters
        in an $S_3$ symmetric Higgs potential. Using these, we probe high-scale behaviour of the scalar
potential. That is, we
        evolve the scalar quartic couplings and require that the model remains
perturbative and keeps
        vacuum stability intact at each intermediate energy scale. Through this exercise, we try to identify
        the parameter space at the input scale that keeps the model \emph{valid}
till very high scales.

    \item Electroweak Symmetry Breaking (EWSB) is triggered when one or
more doublet receives a
    vacuum expectation value (vev). 
While several such configurations
    of the vevs can be there in principle, we consider two such cases
which not only are more relevant
    from the phenomenological point of view, but also demonstrative of the
high-scale validity of the
    $S_3$ potential. For instance, we analyse a 'two inert doublet' scenario
where only one doublet gets
    a vev, and predicts existence of stable scalars through some remnant
symmetry. This scenario thus stands as a potential canditate for describing DM.

    \item The parameter space allowing for high-scale validity is also
subject to various \emph{low
    energy} constraints, i.e., the ones originating from the oblique S, T
and U parameters, signal strength measurements for the 125 GeV Higgs, and also DM searches.

\end{itemize}

This paper is organized as follows. In Section~\ref{Model}, we briefly discuss the salient features of the model, particularly the scalar and Yukawa sectors. The various constraints taken are listed in Section~\ref{Constraints}. The numerical results so obtained are detailed in Section~\ref{Results}, and finally, we summarize in Section~\ref{Conclusions}. Relevant expressions and equations can be found in the Appendix~\ref{Appendix}.

\section{The $S_3$ symmetric three-Higgs-doublet model ($S_3$HDM) in brief.}\label{Model}

\subsection{Scalar sector.}\label{scalar}

The scalar sector consists of three scalar doublets $\phi_1$, $\phi_2$ and $\phi_3$. The most general renormalizable scalar potential consistent with the gauge and $S_3$ symmetries can be cast as~\cite{Aranda:2013kq},  
\begin{subequations}
\begin{eqnarray}
V(\phi)&=& \m_{11}^2(\phi_1^\dagger\phi_1+\phi_2^\dagger\phi_2)+ \m_{33}^2\phi_3^\dagger\phi_3 \nonumber \\
&& + \lambda_1 (\phi_1^\dagger\phi_1+\phi_2^\dagger\phi_2)^2 +\lambda_2 (\phi_1^\dagger\phi_2 -\phi_2^\dagger\phi_1)^2 +\lambda_3 \left\{(\phi_1^\dagger\phi_2+\phi_2^\dagger\phi_1)^2 +(\phi_1^\dagger\phi_1-\phi_2^\dagger\phi_2) ^2\right\} \nonumber \\
&& +\lambda_4 \left\{(\phi_3^\dagger\phi_1)(\phi_1^\dagger\phi_2+\phi_2^\dagger\phi_1) +(\phi_3^\dagger\phi_2)(\phi_1^\dagger\phi_1-\phi_2^\dagger\phi_2) + {\rm h. c.}\right\} \nonumber \\
&& +\lambda_5(\phi_3^\dagger\phi_3)(\phi_1^\dagger\phi_1+\phi_2^\dagger\phi_2) + \lambda_6 \left\{(\phi_3^\dagger\phi_1)(\phi_1^\dagger\phi_3)+(\phi_3^\dagger\phi_2)(\phi_2^\dagger\phi_3)\right\} \nonumber \\
&& +\lambda_7 \left\{(\phi_3^\dagger\phi_1)(\phi_3^\dagger\phi_1) + (\phi_3^\dagger\phi_2)(\phi_3^\dagger\phi_2) +{\rm h. c.}\right\} +\lambda_8(\phi_3^\dagger\phi_3)^2 \,.
\label{quartic}
\end{eqnarray}
\label{potential}
\end{subequations}

A 3HDM is usually known to have CP violating phases \cite{Ivanov:2014doa} in the scalar sector. 
For example a complex $\lambda_{4}$ and $\lambda_7$ in this case leads to CP non-conservation,
although the phases are severely constrained by measurements of Electric Dipole Moment of the 
Neutron (EDMN)\cite{Serebrov:2013tba}. The high-scale stability of a 2HDM is found intact regardless  of the CP phase\cite{Chakrabarty:2014aya}. Thus, the overall conclusions 
regarding validity of the $S_3$HDM at high scales is expected to remain unaffected by the introduction
of CP phases. So we choose $\lambda_{4}$ and $\lambda_7$ to be real henceforth.  

Electro-Weak Symmetry Breaking (EWSB) assigns vacuum expectation values (vevs) $v_1$, $v_2$ and $v_3$ to the doublets $\phi_1$, $\phi_2$ and $\phi_3$ respectively. However, they all are not independent as the $S_3$ invariance forces relationships among them through the minimization conditions below,
\begin{subequations}
\begin{eqnarray}
2\m_{11}^2 &=& -2\lambda_1(v_1^2+v_2^2)-2\lambda_3(v_1^2+v_2^2)-v_3\{6\lambda_4v_2 +(\lambda_5+\lambda_6+2\lambda_7)v_3\} \,, \label{mu11}\\
2\m_{11}^2 &=& -2\lambda_1(v_1^2+v_2^2)-2\lambda_3(v_1^2+v_2^2) -\frac{3v_3}{v_2}\lambda_4(v_1^2-v_2^2) - (\lambda_5+\lambda_6+2\lambda_7)v_3^2 \,, \label{mu22}\\
2\m_{33}^2 &=& \lambda_4\frac{v_2}{v_3}(v_2^2-v_1^2) -(\lambda_5+\lambda_6+2\lambda_7)(v_1^2+v_2^2)-2\lambda_8 v_3^2 \,.
%\label
\end{eqnarray}
\label{minimization}
\end{subequations}
The self-consistency of Eqs.~(\ref{mu11}) and (\ref{mu22}) gives rise to the following possibilities,
\begin{subequations}
\begin{eqnarray}
\lambda_4 &=& 0 \,, \\ \label{faltu} 
{\rm or,}~~v_1 &=& \sqrt{3}v_2 \,, \label{scenarioA} \\
{\rm or,}~~v_1 &=& v_2 = 0, ~v_3 = 246 ~\rm GeV\,. \label{scenarioB}
%\label{}
\end{eqnarray}
\label{consistency}
\end{subequations}
The first case causes a physical scalar to turn massless as reported in \cite{1742-6596-171-1-012028}. This directs us towards the other two cases that we outline below.

The doublets are parameterized in the following fashion,
\be
\phi_{i} = \frac{1}{\sqrt{2}} \begin{pmatrix}
\sqrt{2} w_i^{+} \\
v_i + h_i + i z_i
\end{pmatrix}~ \rm{for}~\textit{i} = 1, 2, 3.
\label{e:doublet}
\ee

The physical scalar spectrum of a generic CP-conserving 3HDM consists of three CP even neutral 
scalars, $H_1$, $H_2$ and $h$; two CP-odd neutral scalars $A_1$ and $A_2$; and two charged
scalars $H_1^+$ and $H_2^+$. We define tan$\beta$ = $\frac{2 v_2}{v_1}$ for Scenario A~(Eqn.\ref{faltu}). For this vev-alignment, 
only two mixing angles $\alpha$ and $\beta$ are sufficient to parameterize the transformation
matrices connecting the SU(2) eigenbasis to the physical basis, somewhat resembling a 2HDM. \footnote{A more detailed discussion
regarding the transformation matrices can be found in \cite{Das:2014fea}.} The model is more conveniently 
described in terms of physical quantities like masses and mixing angles. 
The eight $\l_i$ can be traded for the seven masses and the mixing angle $\a$ (See \cite{Das:2014fea} for definition.) using the following equations,
\begin{subequations}
\begin{eqnarray}
\lambda_1 &=& \frac{1}{2v^2\sin^2\beta} \left\{\left(m_h^2\cos^2\alpha+m_{H_1}^2\sin^2\alpha\right)+ \left(m_{H^+_1}^2 -m_{H^+_2}^2\cos^2\beta -\frac{1}{9} m_{H_2}^2 \right) \right\} \,, \\
\lambda_2 &=& \frac{1}{2v^2\sin^2\beta} \left\{(m_{H^+_1}^2-m_{A_1}^2)- (m_{H^+_2}^2-m_{A_2}^2)\cos^2\beta \right\} \,, \\
\lambda_3 &=& \frac{1}{2v^2\sin^2\beta} \left(\frac{4}{9} m_{H_2}^2 +m_{H^+_2}^2\cos^2\beta - m_{H^+_1}^2\right) \,, \\
\lambda_4 &=& -\frac{2}{9} \frac{m_{H_2}^2}{v^2}\frac{1}{\sin\beta\cos\beta} \,, \\
\lambda_5 &=& \frac{1}{v^2} \left\{\frac{\sin\alpha\cos\alpha}{\sin\beta\cos\beta}\left(m_{H_1}^2-m_h^2 \right) +2 m_{H^+_2}^2 +\frac{1}{9}\frac{m_{H_2}^2}{\cos^2\beta} \right\} \,, \\
\lambda_6 &=& \frac{1}{v^2}\left(\frac{1}{9}\frac{m_{H_2}^2}{\cos^2\beta}+m_{A_2}^2-2m_{H^+_2}^2 \right) \,, \\
\lambda_7 &=& \frac{1}{2v^2}\left(\frac{1}{9}\frac{m_{H_2}^2}{\cos^2\beta}-m_{A_2}^2 \right) \,, \\
\lambda_8 &=& \frac{1}{2v^2\cos^2\beta}\left\{\left(m_h^2\sin^2\alpha+m_{H_1}^2\cos^2\alpha \right) -\frac{1}{9} m_{H_2}^2\tan^2\beta \right\} \,.
%\label
\end{eqnarray}
\label{lambda}
\end{subequations}

We also put forth the Scenario B~(Eqn.\ref{scenarioB}) as an alternate symmetry breaking pattern. In this case, $\phi_3$ is the only active doublet, which is in fact a singlet under $S_3$. Consequently, all the fermions couple to $\phi_3$ alone and thus they too are $S_3$-singlets. The remaining doublets $\phi_1$ and $\phi_2$ remain 
\emph{inert}. A $Z_2$ symmetry is found unbroken for $\l_4$ = 0, and it forbids mixing among scalars coming from different doublets thus enabling one to express the doublets directly in terms of the physical fields as,
\be
\phi_{3} = \frac{1}{\sqrt{2}} \begin{pmatrix}
\sqrt{2} w^{+} \\
v + h + i z
\end{pmatrix}~ 
\label{e:doublet}
\ee
\be
\phi_{i} = \frac{1}{\sqrt{2}} \begin{pmatrix}
\sqrt{2} H_i^{+} \\
H_i + i A_i
\end{pmatrix}~ \rm{for}~\textit{i} = 1, 2.
\label{e:doublet}
\ee
With $m^2_h$ = $2 \l_8 v^2$ now, the $S_3$ symmetry leads to a mass degeneracy in the inert sector,
\bea
m^2_{H_1} &=& m^2_{H_2} = \mu^2_{11} + \frac{1}{2}\l_L v^2 \\
m^2_{A_1} &=& m^2_{A_2} = \mu^2_{11} + \frac{1}{2}\l_A v^2 \\
m^2_{H^+_1} &=& m^2_{H^+_2} = \mu^2_{11} + \frac{1}{2}\l_5 v^2
\eea
Here $-\l_L v$ and $-\l_A v$ denote the $H_1$-$H_1$-$h$ and $A_1$-$A_1$-$h$ couplings respectively. This mass degeneracy can be lifted in this case,
for example, by introducing an $S_3$ breaking quadratic term of the form $-\m_{12}^2(\phi_1^\dagger\phi_2+\phi_2^\dagger\phi_1)$\footnote{The degeneracy persists even after one-loop radiative effects are incorporated. This is because the $Z_2$ symmetry that emerges unbroken
after EWSB is an exact symmetry not only of the scalar potential, but of the entire lagrangian. Thus,
this not only leads to equal 
tree level masses, but also equal couplings for $H_1$ and $H_2$. The two-point correlators for $H_1$ and $H_2$, $\Pi_{H_1 H_1}(p)$ and $\Pi_{H_2 H_2}(p)$ (say) respectively, would have exactly the same expressions then. This would lead to equal one-loop corrected masses for $H_1$ and $H_2$. In other words, the unbroken $Z_2$ symmetry would protect the degeneracy at the one-loop level.}. However, implications of a broken $S_3$ symmetry are outside the scope of this paper.

It is interesting to probe the parameter space arising out of such a vev alignment by proposing $H_1$ and $H_2$ as possible DM canditates. For the $S_3$HDM to qualify as a good DM model, its predictions of relic-density and direct detection rates must be matched against corresponding experimental data. We arrange for the hierarchy
$m_{H_1}$ $\textless$ $m_{A_1}$, $m_{H^+_1}$ throughout our numerical analysis ($H_1$ and $A_1$ are similar
to each other in terms of the masses and couplings, the only difference being the sign
of $\l_7$. Thus a flip in the sign of $\l_7$ would tantamount to interchanging $H_1$ and $A_1$. In that case, $A_1$ would be the DM candiate and the
hierarchy required would be $m_{A_1}$ $\leq$ $m_{H_1},m_{H^+_1}$. The overall physics thus remains unchanged.). LEP constraints 
on the direct search for charged and pseudoscalar Higgs bosons are evaded by taking $m_{H^+_i}$ and $m_{A_i}$ $\textgreater$ 100 GeV\cite{Searches:2001ac}. Similar to the previous case, we describe the model parameter space in terms of the physical parameters $\{$ $\l_1$, $\l_2$, $\l_3$, $m_{H_1}, m_{A_1}, m_{H^+_1}$, $\l_L$ $\}$.

Our main motivation is to study the high-scale stability of the $S_3$HDM for the two different vev assignments discussed above. In doing that we juxtapose the constraints coming from oblique parameters, Higgs signal strengths in the first case, and also the ones coming from relic-density and 
direct detection in the second case. In principle there can be other such vev configurations as well, and our choice is not exhaustive in that sense. Nonetheless, this paper takes into account two representative cases. The first one defines an \emph{active} 3HDM scenario, i.e, when all three $\phi_1$, $\phi_2$ and $\phi_3$ receive non-zero vevs. The second one describes an \emph{inert} scenario, where these inert scalars do not mix with the 125 GeV Higgs that comes from $\phi_3$.

\subsection{Yukawa Sector.}\label{Yukawa}

The most general Yukawa lagrangian consistent with the gauge and $S_3$ symmetries, for the
up-type quarks is given by,
\begin{eqnarray}
- \mathscr L_Y^{u} &=&
\null  y_1^{u} \Big( \bar Q_{1} \tilde\phi_3 u_{1R} 
+ \bar Q_2 \tilde\phi_3 u_{2R} \Big)
+ y_2^{u} \Big\{ \Big( \bar Q_{1}\tilde\phi_2 + \bar
Q_2\tilde\phi_{1}\Big) u_{1R} + 
\Big( \bar Q_{1}\tilde\phi_{1} +
\bar Q_2\tilde\phi_2 \Big)u_{2R} \Big\} \nonumber \\*
&& \null + y_3^{u} \bar Q_3\tilde\phi_3u_{3R}
+y_4^{u} \bar Q_3 \Big( \tilde\phi_1 u_{1R} + \tilde\phi_2u_{2R} \Big)
+y_5^{u} \Big( 
\bar Q_1 \tilde\phi_1 + \bar Q_2\tilde\phi_2 \Big)
u_{3R} + {\rm h.c.}
\label{uYuk}
\end{eqnarray}
The lower component of the $SU(2)$ doublets of Higgs multiplets are 
uncharged in the convention we use. A standard abbreviation reads
 $\tilde\phi_i = i \sigma_2 \phi_i^*$.  The Yukawa
couplings of the $d_R$ quarks can be obtained by replacing $u_{iR}$ by
$d_{iR}$, $y_i^u$ by $y_i^d$, and $\tilde\phi_i$ by $\phi_i$ in $\mathscr L_Y^{u}$
and similarly for leptons.  
The Yukawa couplings are in general complex, which can be
responsible for $CP$ violation. More elaborate discussions on $S_3$ symmetric
Yukawa textures can be found in \cite{Das:2015sca,Canales:2013cga,Ma:2013zca,Teshima:2011wg}.

After symmetry breaking, the mass matrix that arises in the up-type
quark sector is the following, (In the {$u,c,t$} basis):
\begin{eqnarray}
{\cal M}_u = \begin{pmatrix}
(y_1^u v_3 + y_2^u v_2)/\sqrt{2} & y_2^u v_1/\sqrt{2} & y_5^u v_1/\sqrt{2} \\
y_2^u v_1/\sqrt{2}  & y_1^u v_3 - y_2^u v_2/\sqrt{2} & y_5^u v_2/\sqrt{2} \\
y_4^u v_1/\sqrt{2}  & y_4^u v_2/\sqrt{2} & y_3^u v_3/\sqrt{2} \\
\end{pmatrix} \,, \qquad {\rm  with~} v_1=\surd3 v_2 \,.
%\label{}
\end{eqnarray}

The texture is of the same form for the down-type quarks and charged leptons. In principle, one can retain 
all the paramaters in the Yukawa matrix and fine-tune them appropriately in order to reproduce the correct fermion masses and mixings.
However that would make the analysis using RG complicated and unwieldy and hence, we look for a simplification.    
Choosing $y_4^u$, $y_5^u$ = 0 brings ${\cal M}_u$ to a 2 $\times$ 2 $\oplus$ 1 $\times$ 1 block-diagonal form. 
The quark masses in the SM can be straightforwardedly reproduced by diagonalising the remaining the 2 $\times$ 2 block and then tuning the parameters appropriately. 
For example, the choice
$y_1^u$ $\textless$ $y_2^u$ $\textless$ $\textless$ $y_3^u$ reproduces the observed up-type 
quark mass hierarchy. The advantage of this choice is that only $y_3^u$ = $\frac{v}{v_3}y^{SM}_t$
gets a value large enough to cast an impact on the RG evolution, where $y^{SM}_t$ is the SM t-quark Yukawa coupling and, all other Yukawa couplings have a negligible bearing. In addition, even if we invoke a non-zero $y_4^u$ and $y_5^u$, the observed quark-mixings will always render them small. 
Exactly this approximation is applied to the bottom quark and lepton sectors also. It is easy
to see that then $y_3^u$: $y_3^b$: $y_3^l$ = $m_t$: $m_b$: $m_{\tau}$, i.e, this particular approximation
scheme preserves the hierarchy of Yukawa couplings observed in the SM. 
Hence we infer that only the $t$-quark can contribute significantly to the beta functions through the parameter $y_3^u$    
and the effect of all other fermions can be safely neglected in this context. Thus effectively
with only one Yukawa into the picture, as far as high-scale stability is concerned, it becomes easier
to throw light on the scalar sector.

In the inert case, all the fermion generations are $S_3$-singlets and hence couple only to $\phi_3$.

\section{Constraints imposed.}\label{Constraints}

Parameter space of the scenario at hand is surveyed throughly by generating random model-points 
in the $\{$tan$\beta,m_{H^+_1},m_{H^+_2},m_{A_1},m_{A_2},m_{H_1},m_{H_2},c_{\beta-\alpha}$ $\}$ basis in scenario A and, $\{\l_1,\l_2,\l_3,m_{H_1},m_{A_1},m_{H^+_1}, \l_L \}$ in scenario B. We discuss below the various theoretical and experimental constraints imposed to shape the results.

\subsection{Theoretical constraints.}
The $S_3$HDM remains a calculable theory if the model parameters fulfil the respective perturbativity constraints, $|\l_i| \leq 4\pi$, $|y_t|, |g_1|, |g_2|, |g_3| \leq \sqrt{4\pi}$.
A more stringent choice is to demand all of the couplings  $\leq \sqrt{4\pi}$. 
We however stick to 4$\pi$, since this projects out the maximally allowed parameter space.

The 2$\rightarrow$2 amplitude matrix corresponding to scattering of the longitudinal components
of the gauge bosons can be mapped to a corresponding matrix for the scattering of the goldstone 
bosons\cite{PhysRevD.16.1519,Akeroyd:2000wc,Horejsi:2005da,Gorczyca:2011he}. The theory respects unitarity if each eigenvalue of the aforementioned amplitude
matrix does not exceed 8$\pi$.
\begin{eqnarray}
|a_i^\pm|,~|b_i| \le 8\pi, ~\mbox{for}~i=1,2,\ldots,6\,.
\end{eqnarray}
The expressions for the individual eigenvalues\cite{Das:2014fea} in terms of quartic couplings are given below~:
\begin{subequations}
\begin{eqnarray}
a^\pm &=& \left(\lambda_1-\lambda_2+ \frac{\lambda_5+\lambda_6}{2}\right) \pm \sqrt{\left(\lambda_1-\lambda_2+ \frac{\lambda_5+\lambda_6}{2}\right)^2 -4\left\{(\lambda_1-\lambda_2)\left(\frac{\lambda_5+\lambda_6}{2}\right) -\lambda_4^2 \right\} } \,, \\
b^\pm &=& \left(\lambda_1+\lambda_2+2\lambda_3+\lambda_8\right) \pm \sqrt{\left(\lambda_1+\lambda_2+2\lambda_3+\lambda_8\right)^2 -4\left\{\lambda_8(\lambda_1+\lambda_2+2\lambda_3) -2\lambda_7^2 \right\} } \,, \\
c^\pm &=& \left(\lambda_1-\lambda_2+2\lambda_3+\lambda_8\right) \pm \sqrt{\left(\lambda_1-\lambda_2+2\lambda_3+\lambda_8\right)^2 -4\left\{\lambda_8(\lambda_1-\lambda_2+2\lambda_3) -\frac{\lambda_6^2}{2} \right\} } \,, \\
d^\pm &=& \left(\lambda_1+\lambda_2+\frac{\lambda_5}{2}+\lambda_7\right) \pm  \sqrt{\left(\lambda_1+\lambda_2+\frac{\lambda_5}{2}+\lambda_7\right)^2 -4\left\{(\lambda_1+\lambda_2)\left(\frac{\lambda_5}{2}+\lambda_7\right) -\lambda_4^2 \right\} } \,, \\
e^\pm &=& \left(5\lambda_1-\lambda_2+2\lambda_3+3\lambda_8\right) \nonumber \\
&& \pm \sqrt{\left(5\lambda_1-\lambda_2+2\lambda_3+3\lambda_8\right)^2 -4\left\{3\lambda_8(5\lambda_1-\lambda_2+2\lambda_3) -\frac{1}{2}(2\lambda_5+\lambda_6)^2 \right\} } \,, \\
f^\pm &=& \left(\lambda_1+\lambda_2+4\lambda_3+\frac{\lambda_5}{2}+\lambda_6+3\lambda_7\right) \nonumber \\
&& \pm \sqrt{\left(\lambda_1+\lambda_2+4\lambda_3+\frac{\lambda_5}{2}+\lambda_6+3\lambda_7\right)^2 -4\left\{(\lambda_1+\lambda_2+4\lambda_3)\left(\frac{\lambda_5}{2}+\lambda_6+3\lambda_7\right) -9\lambda_4^2 \right\} } \,, \\
h_1 &=& \lambda_5+2\lambda_6-6\lambda_7 \,, \\
h_2 &=& \lambda_5 -2\lambda_7 \,, \\
h_3 &=& 2(\lambda_1-5\lambda_2-2\lambda_3) \,, \\
h_4 &=& 2(\lambda_1-\lambda_2-2\lambda_3) \,, \\
h_5 &=& 2(\l_1+\l_2-2\l_3) \,, \\
h_6 &=& \lambda_5-\lambda_6\,.
%\label
\end{eqnarray}
\label{unitarity eq}
\end{subequations}
In addition to the above, the scalar potential must be bounded from below in order to render the electroweak vacuum stable. Demanding absolute \emph{stability} of the vacuum leads to the following 
conditions\cite{Das:2014fea},
\begin{eqnarray}
\rm vsc1:\l_1 &>& 0 \,, \\
\rm vsc2:\l_8 &>& 0 \,, \\
\rm vsc3:\l_1+\l_3 &>& 0 \,, \\
\rm vsc4:2\l_1 +(\l_3-\l_2) &>& |\l_2+\l_3| \,, \\
\rm vsc5:\l_5 +2\sqrt{\l_8(\l_1+\l_3)} &>& 0 \,, \\
\rm vsc6:\l_5+\l_6+ 2\sqrt{\l_8(\l_1+\l_3)} &>& 2|\l_7| \,, \\
\rm vsc7:\l_1+\l_3+\l_5+\l_6+2\l_7+\l_8 &>& 2|\l_4| \,.
%\label{}
\end{eqnarray}
\label{stability}
This conditions can be arrived at by demanding the scalar potential remains positive along
various directions in the field space in the $\phi_i \rightarrow \infty$ limit. We do not consider 
\emph{metastable} vacua configurations in this paper~\cite{Branchina:2013jra,PhysRevD.91.013003}, which are expected to project a more relaxed parameter space in this context.

\subsection{Oblique parameters.} 
%\textbf{(To include or not to include, this subsection?)}\\
The $S_3$HDM induces modification in the $S$, $T$ and $U$ parameters through the additional scalars participating in the loops of the gauge-boson self energies. One discerns the 3HDM contribution from the SM as,
\besub
\bea
S = S_{SM} + \Delta S \\
T = T_{SM} + \Delta T,\\
U = U_{SM} + \Delta U
\eea
\eesub
Here $\Delta S$, $\Delta T$ and $\Delta U$ denote the $S_3$HDM contributions. These have been derived following the approach outlined in\cite{Grimus:2008nb}. Relevant expressions to can be found in the Appendix~\ref{ss:RGE}. 
The central value is the contribution coming from the standard model with the reference values $m_{h,\rm{ref}} = 125.0$ GeV and $M_{t,\rm{ref}} = 173.1$ GeV where $M_t$ denotes the pole mass of the top quark. We have used 1$\sigma$ limits of $S$,$T$ and $U$ following~\cite{Baak:2013ppa}.

\subsection{ Higgs Signal-strengths}
 The  ATLAS and CMS collaborations have measured the production cross section
for a $\sim$125 GeV Higgs multiplied by its branching ratios
to various possible channels. The results so far are increasingly in favor of the SM predictions.
An extended Higgs sector, such as the $S_3$HDM although can very well contain a scalar 
with mass around 125 GeV, 
but yet can potentially modify the signal strength predictions through its modified higgs-gauge boson
and higgs-fermion couplings. For example, the $hVV$ and $hb\bar{b}$ couplings get scaled by  sin($\beta-\alpha$) and $\frac{sin\alpha}{cos\beta}$ w.r.t the SM, in the 
case with three non-zero vevs. The loop induced decay widths to $\gamma\gamma$ and $Z\gamma$ final states are also modified. However, one can always arrange
for $\a = \b - \frac{\pi}{2}$ which reproduces exact SM couplings. This so called \emph{alignment limit} is present in two-higgs doublet models as well\cite{Gunion:2002zf}. We explore a case where this limit is not strictly enforced, rather sin$(\b - \a)$ = 0.98 is taken. The reader is reminded that the tree level couplings of $h$ to fermions and gauge bosons remain identical to the SM in the 
presence of additional inert doublets.

In order to check the consistency of a 2HDM with the measured rates in various channels,
we theoretically compute the signal strength $\mu^{i}$ for the $i$-th channel using the relation:
\be
\mu^{i} = \frac{R_{\rm{prod}} \times R_{\rm{decay}}^i}{R_{\rm{width}}}~.
\ee    
Here $R_{\rm{prod}}$, $R_{\rm{decay}}^i$ and $R_{\rm{width}}$ denote respectively the ratios of the theoretically calculated production cross section, the decay rate to the $i$-th channel and the total decay width for a $\sim$125 GeV Higgs to their corresponding SM counterparts. For our numerical analysis, we have taken gluon fusion to be the dominant production mode for the SM-like Higgs.\footnote{While other channels such as vector boson fusion (VBF) and associated Higgs production with W/Z (VH) have yielded data in the 8 TeV run, the best fit signal strengths are still dominated by the gluon fusion channel.} The predicted signal-strengths to $ZZ$, $WW$, $b\bar{b}$ channels are in excellent agreement with the SM once the alignment limit is invoked. $\mu_{\gamma \gamma}$ still 
needs to be controlled since the charged scalars do not decouple from the theory in spite of an 
exact alignment (see Appendix~\ref{h2gg}). In an exact-alignment scenario, the total width
of $h$ hardly deviates from its SM value and $\mu_{\gamma \gamma}$ settles approximately to $\frac{\Gamma^{S_3 \rm HDM}_{h \rightarrow \gamma \gamma}}{\Gamma^{\rm SM}_{h \rightarrow \gamma \gamma}}$. Latest measurements from ATLAS and CMS give $\mu_{\gamma \gamma}$ = $1.17^{+0.27}_{-0.27}$ and $1.12^{+0.24}_{-0.24}$ respectively~\cite{Aad:2014eha,Khachatryan:2014jba}. We use the cited limits at 2$\sigma$.

We make the passing remark that in-house codes have been employed to carry out the computations related to oblique parameters and signal strengths. In particular, the RG equations have been 
numerically solved by implementing the Runge-Kutta (RK4) algorithm in the same.

\subsection{Dark matter relic density and direct detection}
In the one active + two inert doublet case, we impose that the relic density must be away 
by at most 3$\sigma$ limits from the PLANCK\cite{Ade:2013zuv} central value. That is,
\be
0.1118 \le \Omega h^2 \le 0.1199.
\ee
A more relaxed requirement is to impose only the upper limit, in which case it implies that the $S_3$ inert scalars only partially account for the observed relic density. Relic density calculations in this work are done using the publicly available code \texttt{micrOMEGAs}~\cite{Belanger:2013oya}.

Experiments like XENON100\cite{Aprile:2012nq}, LUX\cite{Akerib:2013tjd} have placed upper limits on WIMP-nucleon scattering
cross sections. We again use \texttt{micrOMEGAs} to compute the cross sections and adhere to the more stringent constraints by LUX. Given that WIMP-nucleon scattering in this model occurs only through a t-channel $h$ exchange, the cross section computation is plagued by the uncertainty in the strange quark form factor. We have resorted to the \texttt{micrOMEGAs} default
parameters in this regard. We have imposed an upper bound of $10^{-46}$ $\rm cm^2$ on the
spin-independent WIMP-nucleon cross section throughout our analysis.

\subsection{Evolution under Renormalisation Group.}

The main motivation of this paper is to study the behaviour of the $S_3$HDM parameters under Renormalisation Group (RG). The \emph{strategy} adopted is, we select parameter points consistent with the constraints discussed above. The parameter space obtained in the process is allowed to evolve under RG. The one loop beta functions employed for this analysis are listed in the appendix.
They were derived by demanding scale-invariance of the one-loop corrected scalar potential following \cite{Ferreira:2009jb}, and, cross checked by a standard Feynman diagrammatic calculation. Constraints stemming from perturbativity, unitarity and vacuum stability are demanded to be fulfilled throughout the course of evolution, up to some cut-off $\L$. There is however, no natural choice for $\L$, given the fact we assume that $S_3$HDM is the only physics up to this scale. We aim to push $\L$ to as high as the GUT scale, or the Planck scale, and explore the consequences on our scenario. Incorporation of these constraints in the RG evolution tightens up the parameter space at the electroweak scale. 

Discussion of the RG constraints is crucial in context of a non-minimal Higgs sector such as the $S_3$HDM, owing to the fact that the additional scalars could ameliorate the vacuum instability problem in the SM\cite{Sher:1988mj}. However, due to the additional bosonic content, quartic couplings tend to rise fast and hit the \emph{Landau pole} even though vacuum stability is preserved. To strike a balance between these extremes, the model parameters have to be judiciously tuned. This is precisely what we aim to do in context of an $S_3$HDM. 

\section{Impact of the constraints on the parameter space.}\label{Results}

\subsection{Scenario A: $v_1$ = $\sqrt{3} v_2$.}

Model points are sampled randomly through a scan of the parameter space within the specified ranges,
\bea
\rm tan\beta \in  [0.1,50] \nonumber \\
m_{H_1}, m_{H_2} \in  [125 ~\rm GeV, 1000 ~\rm GeV] \nonumber \\
m_{A_1}, m_{A_2} \in  [100 ~\rm GeV, 1000 ~\rm GeV]  \nonumber \\
m_{H^+_1}, m_{H^+_2} \in  [80 ~\rm GeV, 1000 ~\rm GeV] \nonumber 
\eea

Demanding perturbativity at the electroweak scale puts upper bounds on the scalar masses and
tan$\beta$. In particular, all the scalar masses lie below $\sim$ 800 GeV and tan$\beta$ $\in$ [0.3,~13.6]. The upper bounds on the masses and tan$\beta$ settle at 1 TeV and 17.3 respectively upon relaxing the perturbativity constraint. In that case the bounds are put by unitarity alone, an observation in consonance with the findings in \cite{Das:2014fea}. Any value of tan$\beta$ 
outside the quoted limit is responsible for making the theory non-perturbative through the large values it gives to the quartic couplings in the process. This can be revealed through an inspection of Eqn.(\ref{lambda}). 

\begin{figure} %[t]
\begin{center}
%\rotatebox{90}{\quad\quad\quad\quad$m_A$ (GeV)}
%
\includegraphics[scale=0.43]{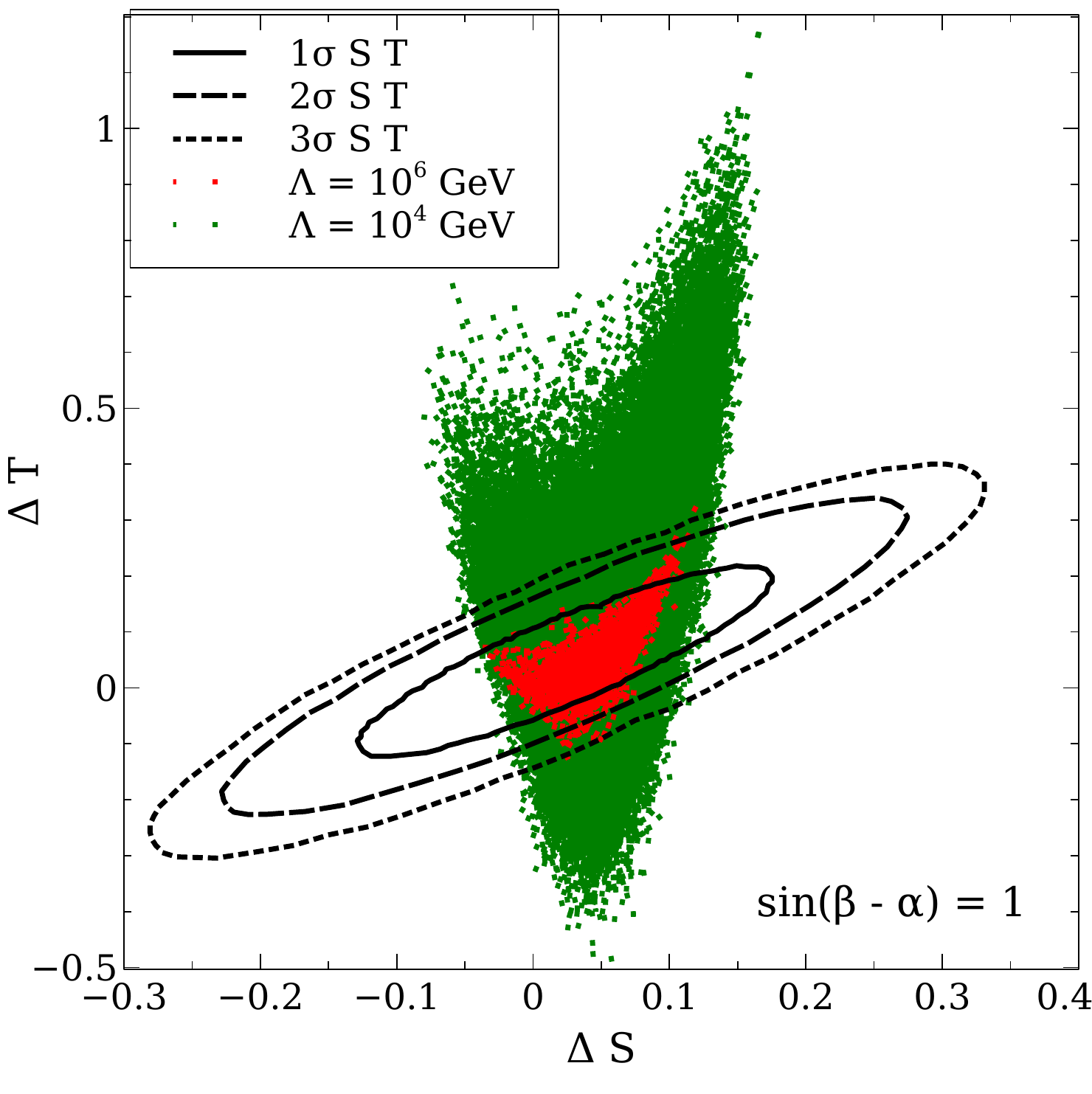}~~~ 
\includegraphics[scale=0.43]{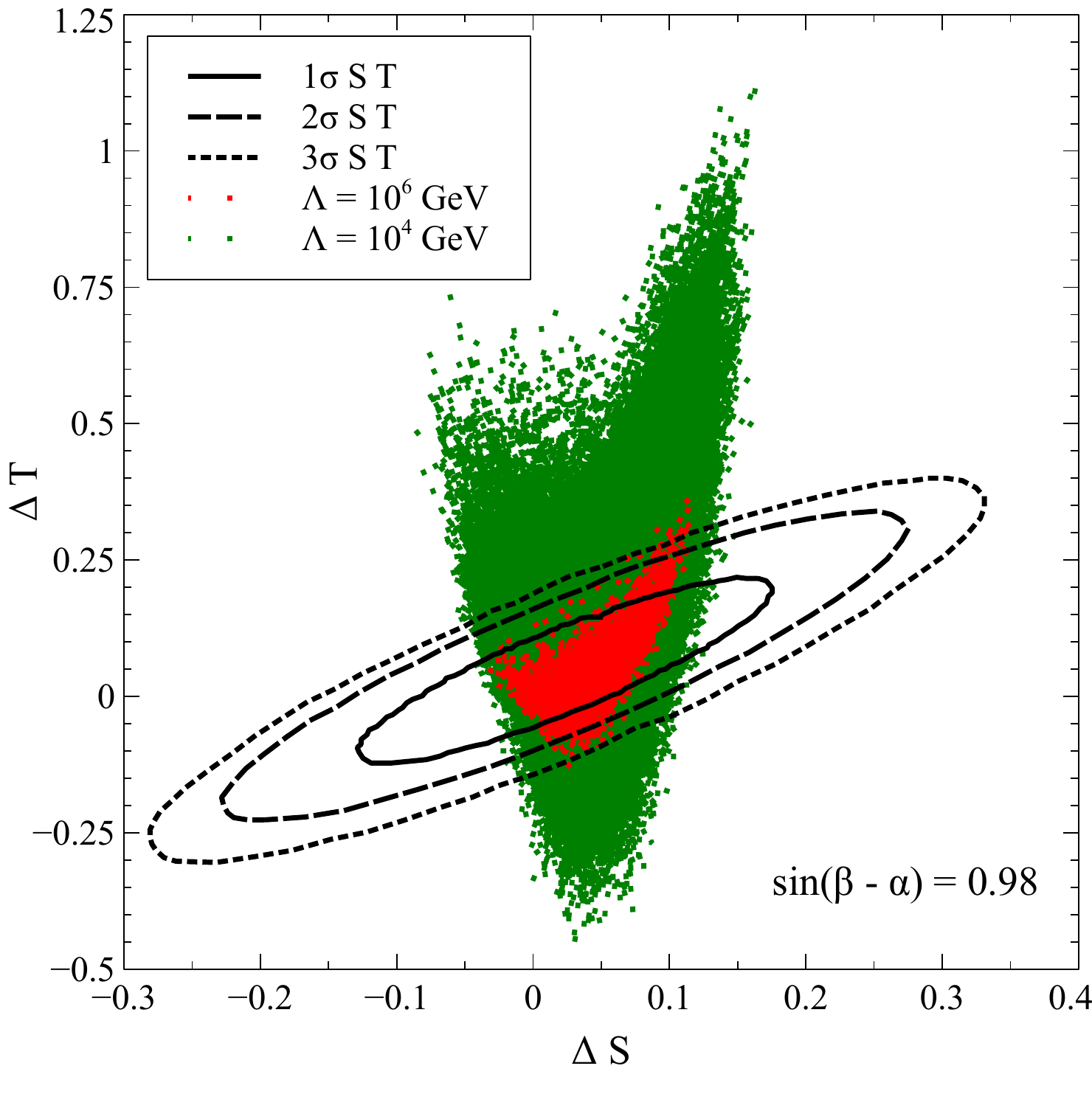} 

\caption{Contribution of the $S_3$HDM scalars to the oblique paramaters for sin($\beta-\alpha$) = 1.0 (Left) and sin($\beta-\alpha$) = 0.98 (Right). The ellipses denote
the 1$\sigma$ (solid), 2$\sigma$ (dashed) and 3$\sigma$ (dotted) limits. The green and red points
indicate validity till $10^4$ GeV and $10^6$ GeV respectively. We notice that the oblique parameters do not change appreciably for a slight departure from exact alignment.}
\label{f:ST}
\end{center}
\end{figure}

The next part of the analysis involves evolution under RG. The key finding here is that this scenario is not valid beyond $10^{7}$ GeV. This as attributed to the following two reasons (i) Quartic couplings are large at the input scale itself, they hit the perturbative limit around the multi-TeV
scale. This can be understood using the following logic, the quartic couplings at the input scale are typically $\sim$ $\frac{m^2}{v^2}$ (see Eqn.(\ref{lambda})), where $m$ refers to any physical $S_3$HDM mass. Thus for
an $m$ below the TeV scale, at least one quartic coupling becomes large enough to make the theory non-perturbative. (ii) tan$\beta \textgreater 3$ in particular destabilises the vacuum by enhancing the t-Yukawa
with respect to its SM value. It so happens that for many parameter points, the Yukawa coupling itself
evolves to non-perturbvative value below the instability scale, however this is a subleading effect. 
The $T$ parameter constraint negates a large number of scan points, many of which otherwise clear the RG constraints up to the highest permissible cut-off $10^7$ GeV. This we show in Fig.(\ref{f:ST}). $\Delta S$ mostly stays within its 1$\sigma$ limit. We also prepare the following two benchmarks models (Table ~\ref{BP1to2}) to reinforce our observation on a violated vacuum stability or unitarity.

\begin{table}[h]
\centering
\begin{tabular}{|c c c c c c c c|}
\hline
Benchmark  & tan$\beta$  & $m_{A_1}$(GeV) & $m_{A_2}$(GeV) & $m_{H^+_1}$(GeV) & $m_{H^+_2}$(GeV) & $m_{H_1}$(GeV) & $m_{H_2}$(GeV)\\ \hline \hline
BP1 & 3.54 & 265.12 & 392.00 & 146.00 & 105.00 & 233.77 & 143.05\\ \hline
BP2 & 1.02 & 102.22 & 167.78 & 119.80 & 107.00 & 214.95 & 132.35\\ \hline
\end{tabular}
\caption{Benchmark points chosen to illustrate the behaviour under RGE. $\L$ denotes the maximum
extrapolation scale up to which vacuum stability and perturbativity are ensured.}
\label{BP1to2}
\end{table}

\begin{figure} %[t]
\begin{center}
%\rotatebox{90}{\quad\quad\quad\quad$m_A$ (GeV)}
%
\includegraphics[scale=0.40]{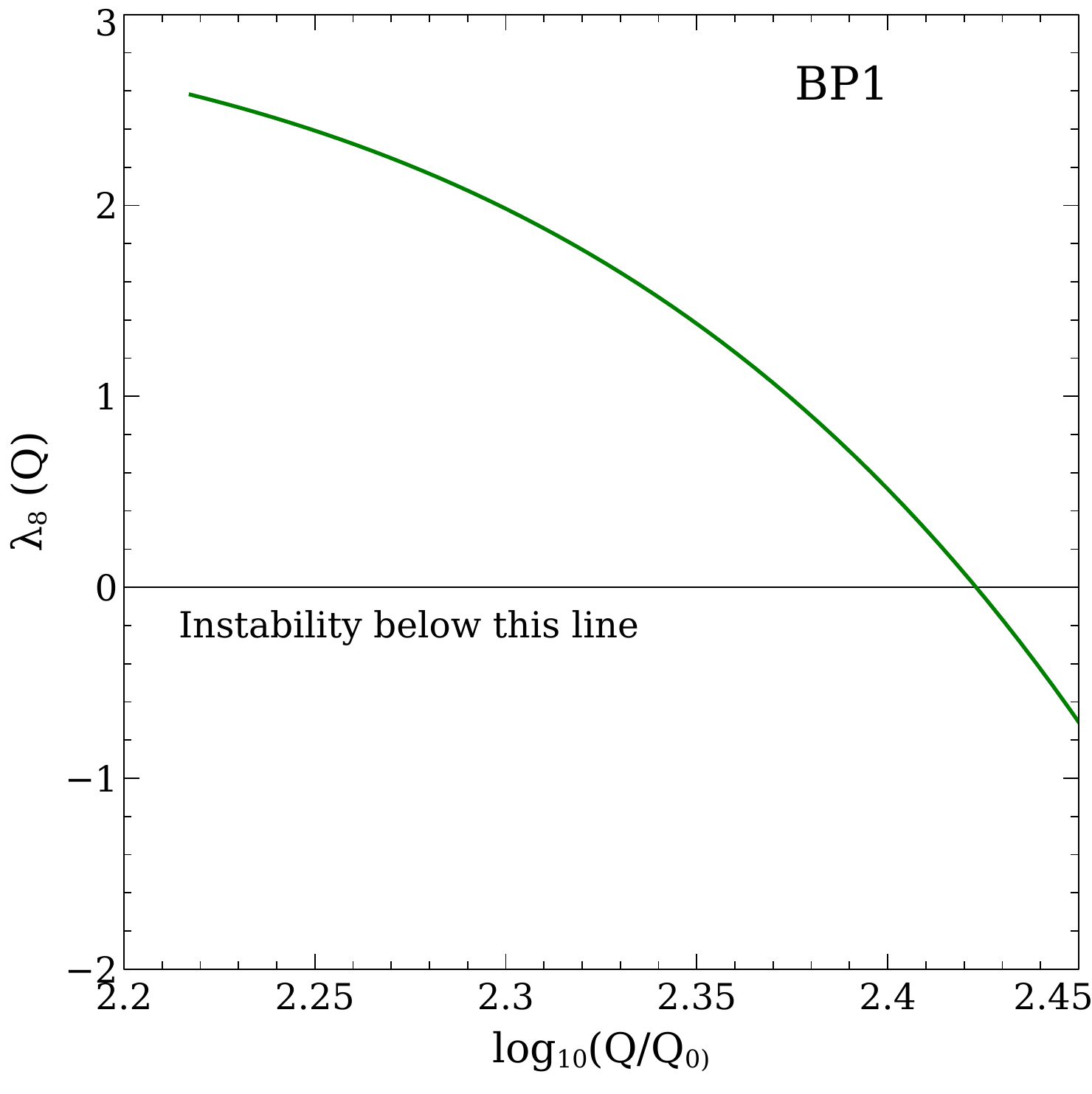}~~~ 
\includegraphics[scale=0.40]{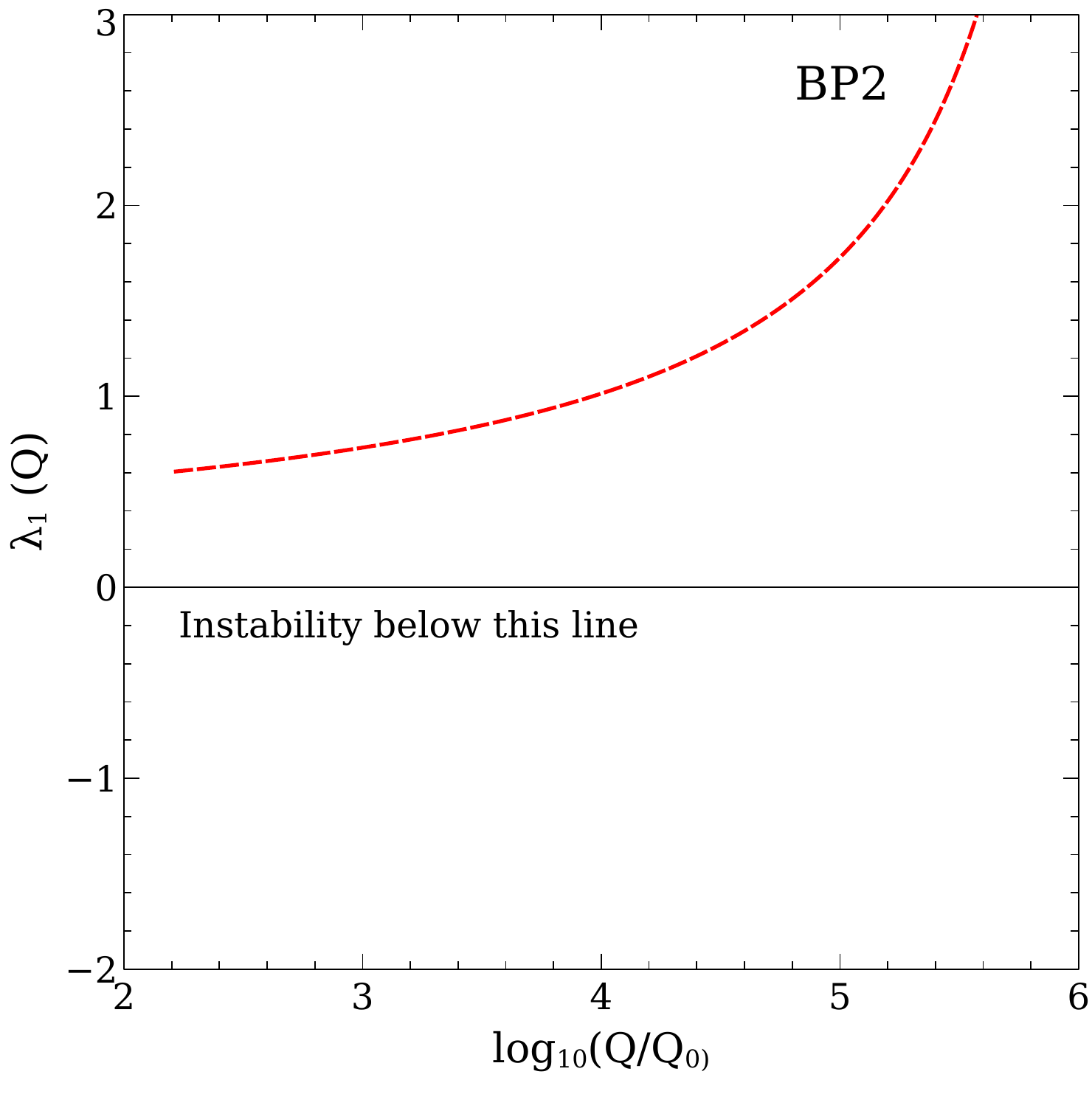}~~~
\caption{Running of $\l_8$ corresponding to BP1 (left) and $\l_1$ corresponding to BP2 (right).
$m_h$ = 125 GeV and an exact alignment sin($\beta-\alpha$) = 1.0 taken in both.}
\label{f:running_active}
\end{center}
\end{figure}

BP1 leads to a destabilised vacuum through $\l_8 \textless 0$ occuring  below the TeV scale. On the
other hand, $\l_1$ in BP2 rises rapidly and quickly becomes non-perturbative just after crossing $10^6$ GeV. The running of $\l_8$ and $\l_1$ in the two cases is displayed in Fig.(\ref{f:running_active}).

The bounds finally obtained on $\l_i$, taking into account the oblique parameter and the diphoton constraints, are are summarised in the Table~\ref{lambda1to8}. 
\begin{table}[h]
\centering
\begin{tabular}{|c c c c|}
\hline
Parameter   & $\L = 10^3$ GeV & $\L = 10^4$ GeV            & $\L = 10^6$ GeV \\ \hline \hline

$\l_1 \in$  & [0, 2.7] & [0, 1.4] & [0, 0.7] \\ \hline
$\l_2 \in$  & [-2.7, 2.5] & [-1.4, 1.3] & [-0.6, 0.6] \\ \hline
$\l_3 \in$  & [-2.2, 2.6] & [-1.0, 1.3] & [-0.2, 0.6] \\ \hline
$\l_4 \in$  & [-2.1, -0.1] & [-0.9, -0.1] & [-0.4, -0.1] \\ \hline
$\l_5 \in$  & [-2.7, 5.5] & [-1.1, 3.0] & [-0.4, 1.5] \\ \hline
$\l_6 \in$  & [-5.3, 4.0] & [-2.6, 1.9] & [-1.1, 0.7] \\ \hline
$\l_7 \in$  & [-2.2, 0.9] & [-1.0, 0.3] & [-0.4, 0] \\ \hline
$\l_8 \in$  & [0, 3.8] & [0, 1.9] & [0.1, 1.1] \\ \hline
\end{tabular}
\caption{Bounds on the quartic couplings, for $\L = 10^3, 10^4, 10^6$ GeV. Oblique parameter
and diphoton constraints are also taken into account. We show the numbers up to the 
first decimal place.}
\label{lambda1to8}
\end{table}

\begin{figure} %[t]
\begin{center}
\includegraphics[scale=0.40]{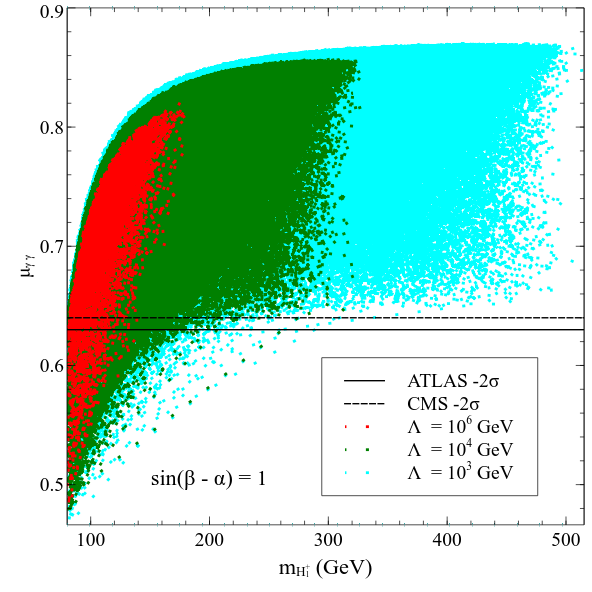}~~~ 
\includegraphics[scale=0.40]{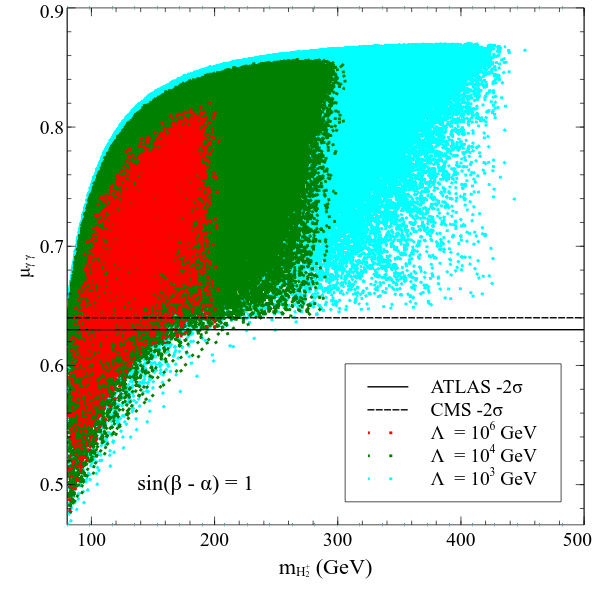}
\includegraphics[scale=0.40]{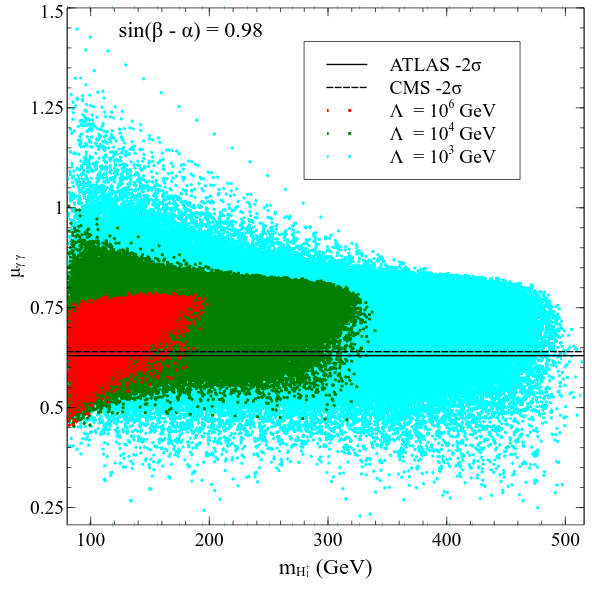}~~~ 
\includegraphics[scale=0.40]{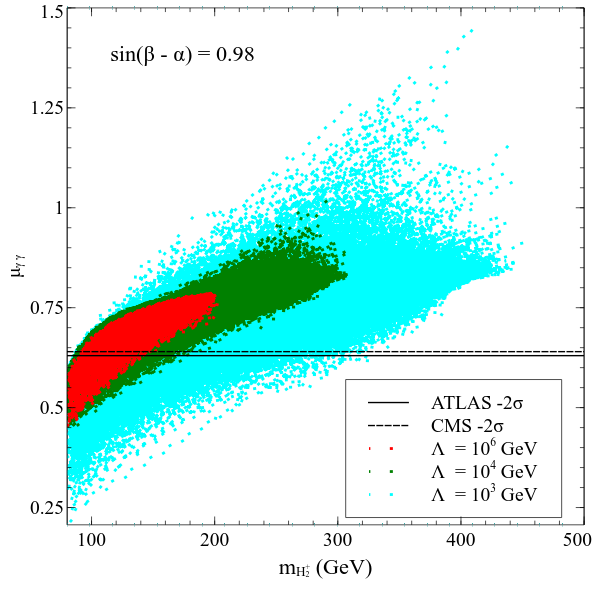}
\caption{$h\rightarrow \gamma \gamma$ rates for an $S_3$HDM valid till a cut-off $\L$. The cyan, green and red points are respectively for $\L = 10^{3}, 10^{4} ~\rm and ~10^{6}$ GeV. The solid and dotted lines denote the 2$\sigma$ limits below the central value given by ATLAS and CMS respectively
. }
\label{f:mugaga}
\end{center}
\end{figure}

\begin{figure} %[t]
\begin{center}
%\rotatebox{90}{\quad\quad\quad\quad$m_A$ (GeV)}
%
\includegraphics[scale=0.38]{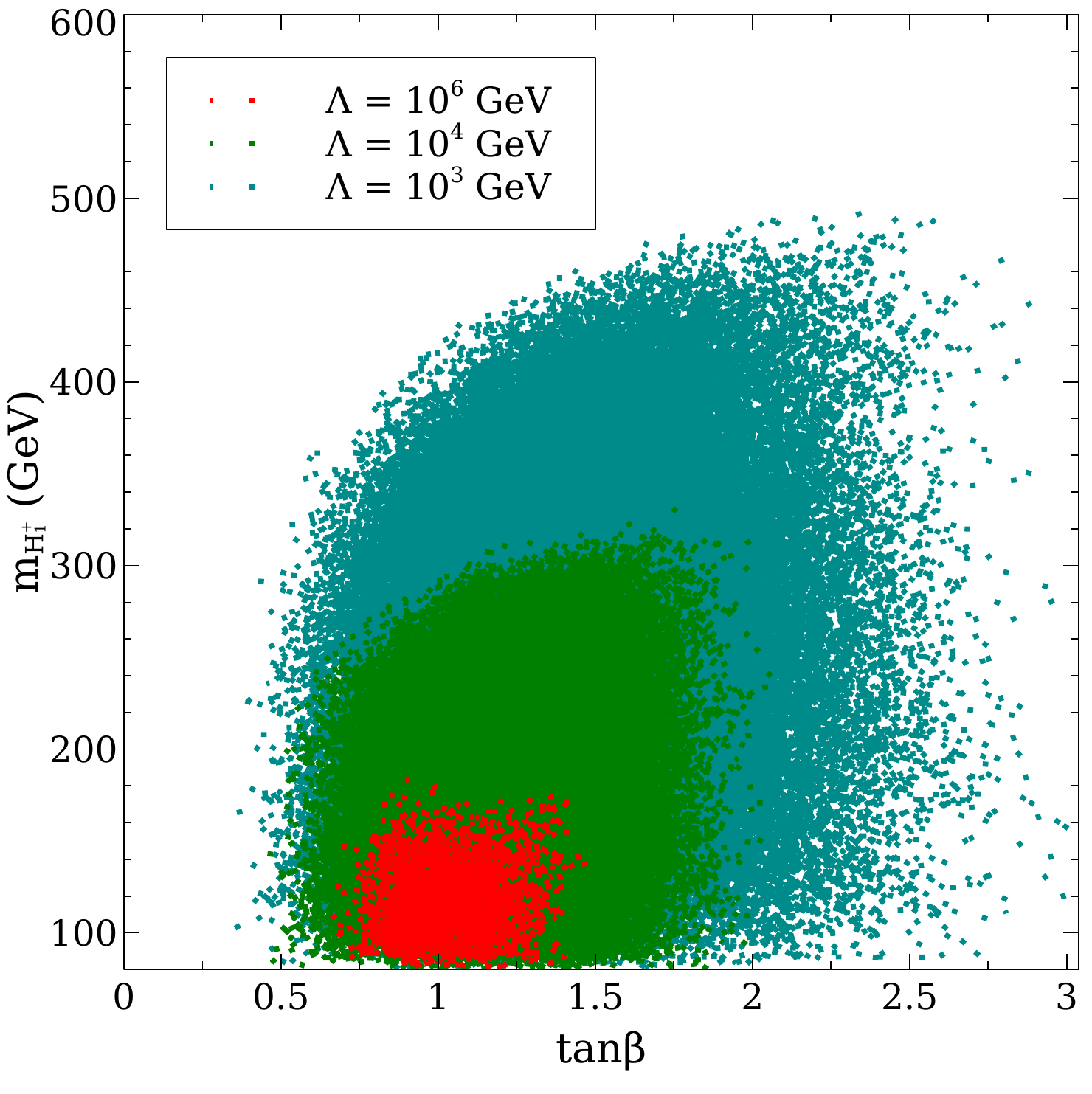}~~~ 
\includegraphics[scale=0.38]{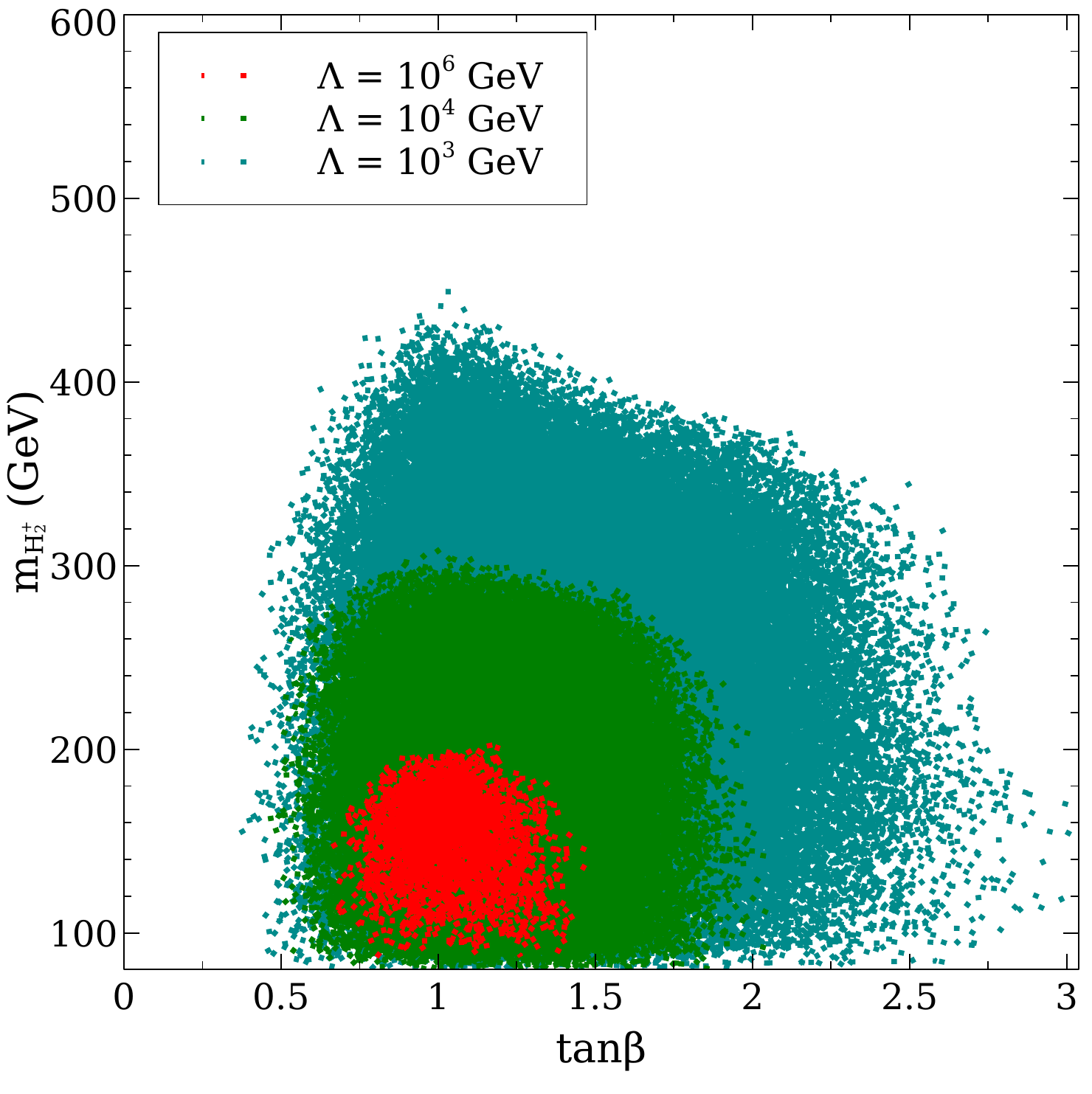}~~~ 
\includegraphics[scale=0.38]{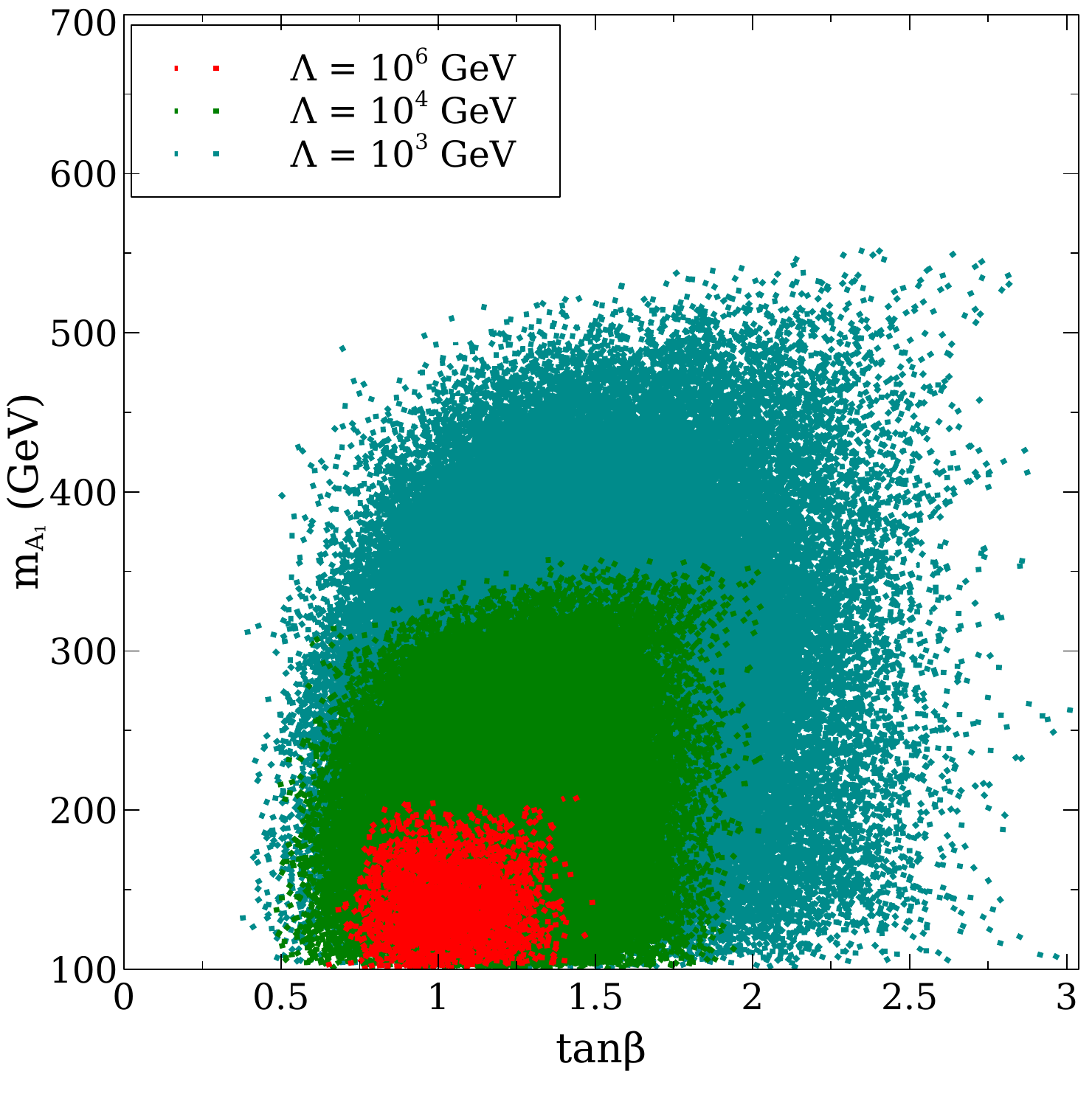}~~~ 
\\
\includegraphics[scale=0.38]{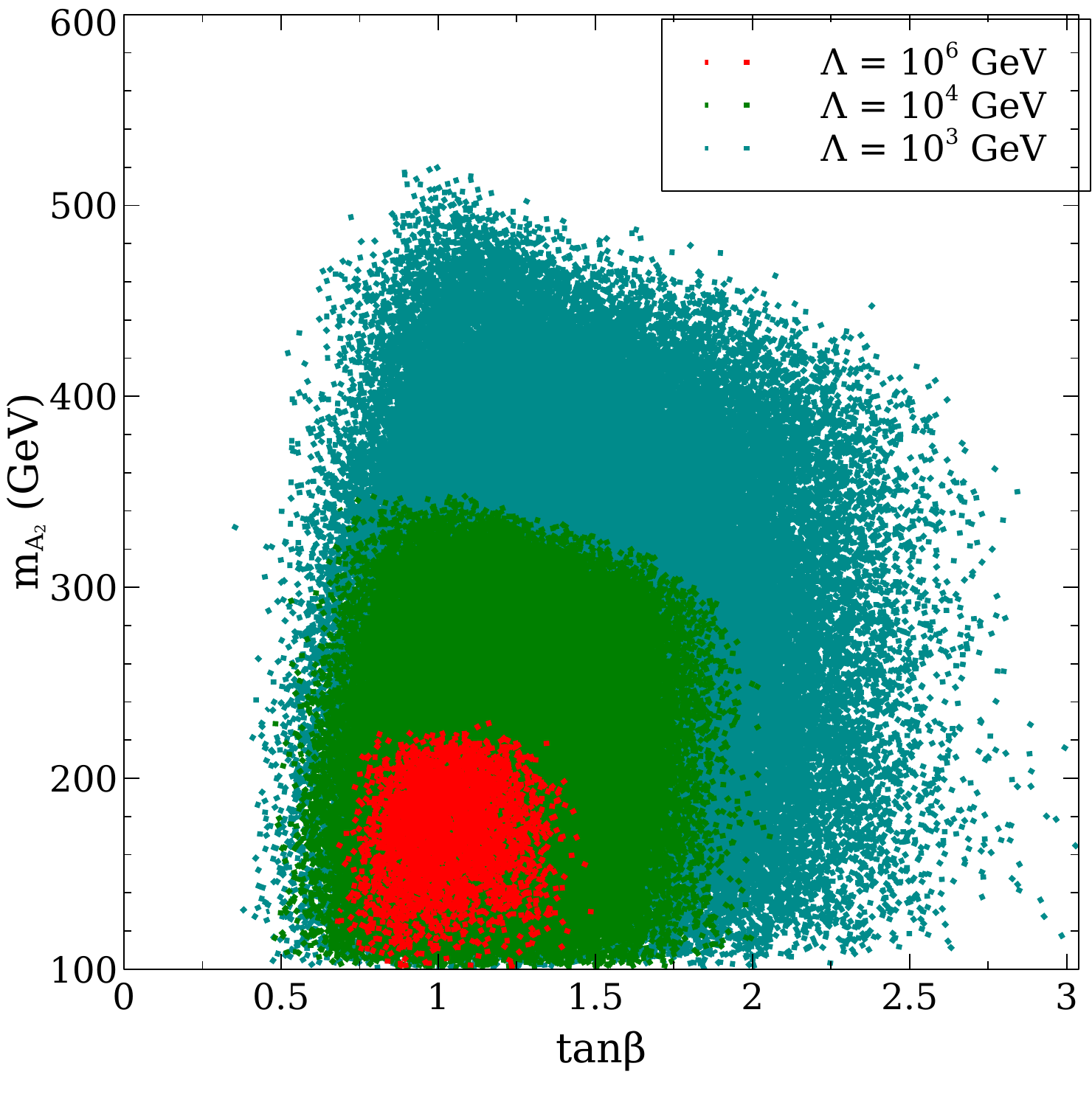}~~~ 
\includegraphics[scale=0.38]{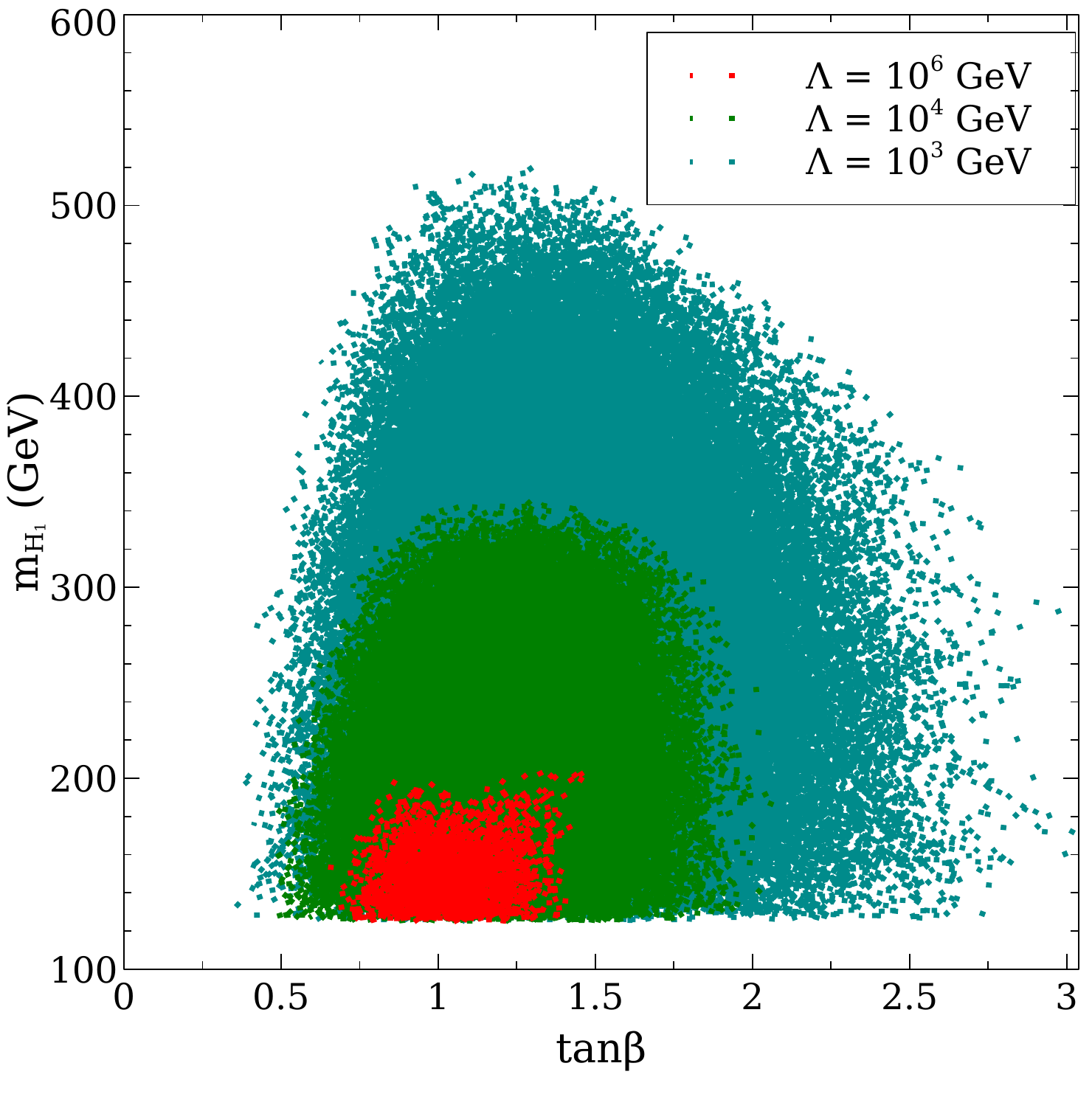}~~~ 
\includegraphics[scale=0.38]{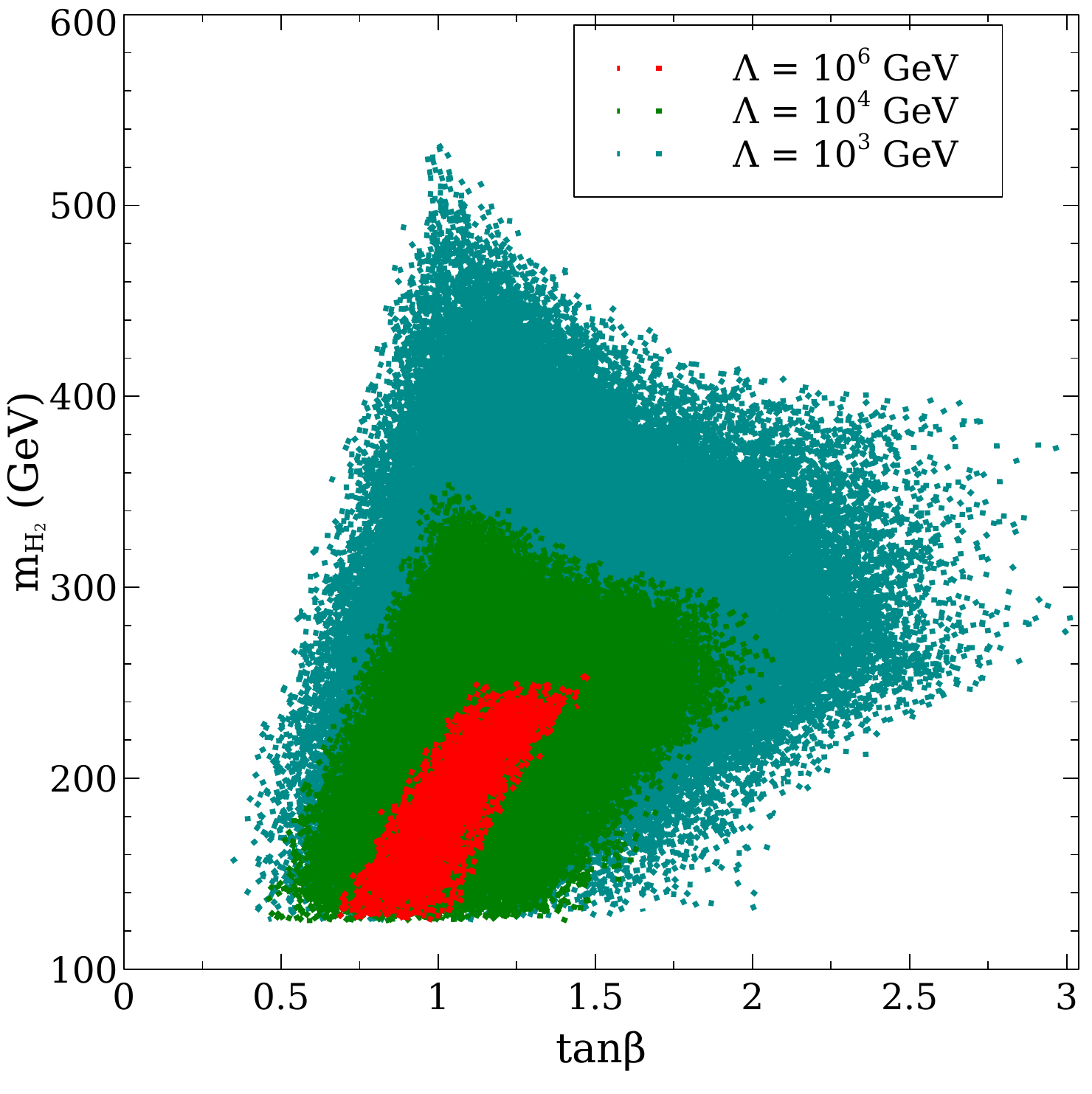}~~~ 
\\
\caption{Regions consistent with the theoretical constraints up to a given cut-off. The cyan, green and red points are valid till $10^3$ GeV, $10^4$ GeV and $10^6$ GeV respectively. Oblique parameter
and diphoton constraints are also taken into account. Points valid till $10^7$ GeV get disallowed
by the diphoton constraint and are hence not displayed. An exact alignment is chosen and it has been checked that the bounds do not change for a small deviation from exact alignment.}
\label{f:masses}
\end{center}
\end{figure}

The $h \rightarrow \gamma \gamma$ rate diminishes with respect to the SM throughout the parameter space, however only for a strict imposition of sin($\b-\a$) = 1\cite{Das:2014fea}. 
We have projected the $S_3$HDM $\mu_{\gamma \gamma}$ values versus $m_{H^+_1}$ and $m_{H^+_2}$ in Fig.(\ref{f:mugaga}). The dimensionful $h H^+_i H^-_i$ denoted by $g_{h H^+_i H^-_i}$ is conveniently
expressed through~ $g_{hH^+_iH^-_i } = \frac{2\kappa_im^2_{H^+_i}}{v}$ , where $\kappa_i$ are dimensionless. Whenever $\alpha = \beta - \frac{\pi}{2}$, it is seen that $\kappa_i = -\Big(1 + \frac{m^2_h}{2 m^2_{H^+_i}}\Big)$ \cite{Das:2014fea} (exact expression given in Appendix \ref{h2gg}). A decrement in
$\mu_{\gamma \gamma}$, in an exact alignment case thus becomes inevitable, since both $\kappa_1$ and $\kappa_2$ are always negative (see Appendix \ref{h2gg}). In fact, $\mu_{\gamma \gamma}$ never exceeds 0.82 for validity till $10^6$ GeV, given that
$|\kappa_1|$, $|\kappa_2|$ $\geq$ 1.39 in that case. Following a similar trend, the points valid till  $\L$ = $10^7$ GeV give $\mu_{\gamma \gamma}$ $\textless$ 0.63 and hence are not phenomenologically acceptable. The bounds put on $\l_i$ translate into corresponding bounds on tan$\beta$ and the non-standard scalar masses, as shown in Fig.\ref{f:masses}. We point out that while $m_{H_2}$ could be up to 270 GeV for $\L$ = $10^6$ GeV, the other masses do not exceed 210 GeV for most parameter points.
It is to be noted that $\kappa_1$ and $\kappa_2$ can take either sign for departure from exact
alignment, and hence an increment in the diphoton rate is possible there. (See Fig.(\ref{f:mugaga}) for sin($\b - \a$) = 0.98.)

A generic feature in context of Scenario A is that, 
the mass parameters $m^2_{11}$ and $m^2_{33}$ get traded off through the tadpole conditions,
making $\l_i$ expressible in terms of the physical scalars only. Thus for physical 
scalars luking below 1TeV, $\l_i$ are already $\mathcal{O}$(1) or even larger at the input scale. 
This does not lead to a model that is perturbative at a high scale. As a possible remedy, additional mass parameters in the equations relating $\l_i$ to the physical masses could induce cancellations 
keeping the quartics further small at the EW scale. This could be achieved either through inclusion 
of quadratic terms violating $S_3$, or through invoking an $\textit{inert}$ vev structure where all of 
$m^2_{11}$ and $m^2_{33}$ do get not eliminated. These terms can elevate the non-standard masses to around $\sim$ 1 TeV and can also lead to $\mu_{\gamma \gamma} \textgreater$ 1. Since a broken $S_3$ group is beyond the ambit of the present study, we focus on the inert case (Scenario B)
in the subsequent section.

\subsection{Scenario B: $v_1$ = $v_2$ = 0, $v_3$ = 246 GeV.}

One needs to put $\l_4$ = 0 in order to keep the DM stable through an unbroken $Z_2$ symmetry. 
Correct relic density is obtained in the mass regimes $m_{H_1} \textless ~\rm ~80 ~GeV$ and $m_{H_1} \textgreater ~\rm 370 ~GeV$. We discuss below the phenomenology in detail. 

\subsubsection{$m_{H_1} \textless ~\rm 80 ~GeV$.}

DM particles dominantly annihilate to the $b\bar{b}$ final state through an $h$ in the s-channel, in this mass regime. A sharp decline in relic abundance is noted for 
$m_{H_1} \textgreater ~\rm 80 ~GeV$, when the $VV$ ($V$ denoting a vector boson.) mode opens up. Maintaining appropriate mass gaps amongst $H_1$, $A_1$ and $H^+_1$ turns advantageous in the two following ways. Firstly, the DM relic abundance does not deplete fast through co-annihilations 
brought in by by a narrow mass splitting. Secondly, it gives sizable values to $\l_5$, $\l_6$ and $\l_7$ which in turn aid to stabilize the vacuum far beyond the SM instability scale, even up to the Planck scale. Overall, the 
phenomenology in this mass regime is broadly similar to the case with a single inert doublet. 

\begin{table}[h]
\centering
\begin{tabular}{|c c c c c c c c|}
\hline
Benchmark  & $m_{H_1}$(GeV) & $m_{A_1}$(GeV) & $m_{H^+_1}$(GeV)  & $\l_L$  & $\Omega h^2$ & $\sigma_{SI} (cm^2)$   & $\L$(GeV)\\ \hline \hline
BP3 & 57.00 & 102.00 & 120.00 & 0.0042 & 0.1170 & $4.63 \times 10^{-47}$ & $10^{19}$\\ \hline
\end{tabular}
\caption{Benchmark point illustrating the behaviour under RGE. $\L$ denotes the maximum
extrapolation scale up to which vacuum stability and perturbativity are ensured.}
\label{BP3}
\end{table}

The displayed benchmark BP3 (Table \ref{BP3}) keeps BR($h\rightarrow \rm invisible$) 
$\textless$ 19 $\%$ owing to the tiny $\l_L$. A perturbative theory at high scales 
requires $m_{A_1}$ and $m_{H^+_1}$ to obey sharp upper bounds, a feature not reflected
by the DM constraints alone. For instance, we need $m_{A_1}$, $m_{H^+_1}$ $\textless$ 135 GeV 
in order to salvage perturbativity till the GUT scale.

\begin{figure} %[t]
\begin{center}
%\rotatebox{90}{\quad\quad\quad\quad$m_A$ (GeV)}
%
\includegraphics[scale=0.45]{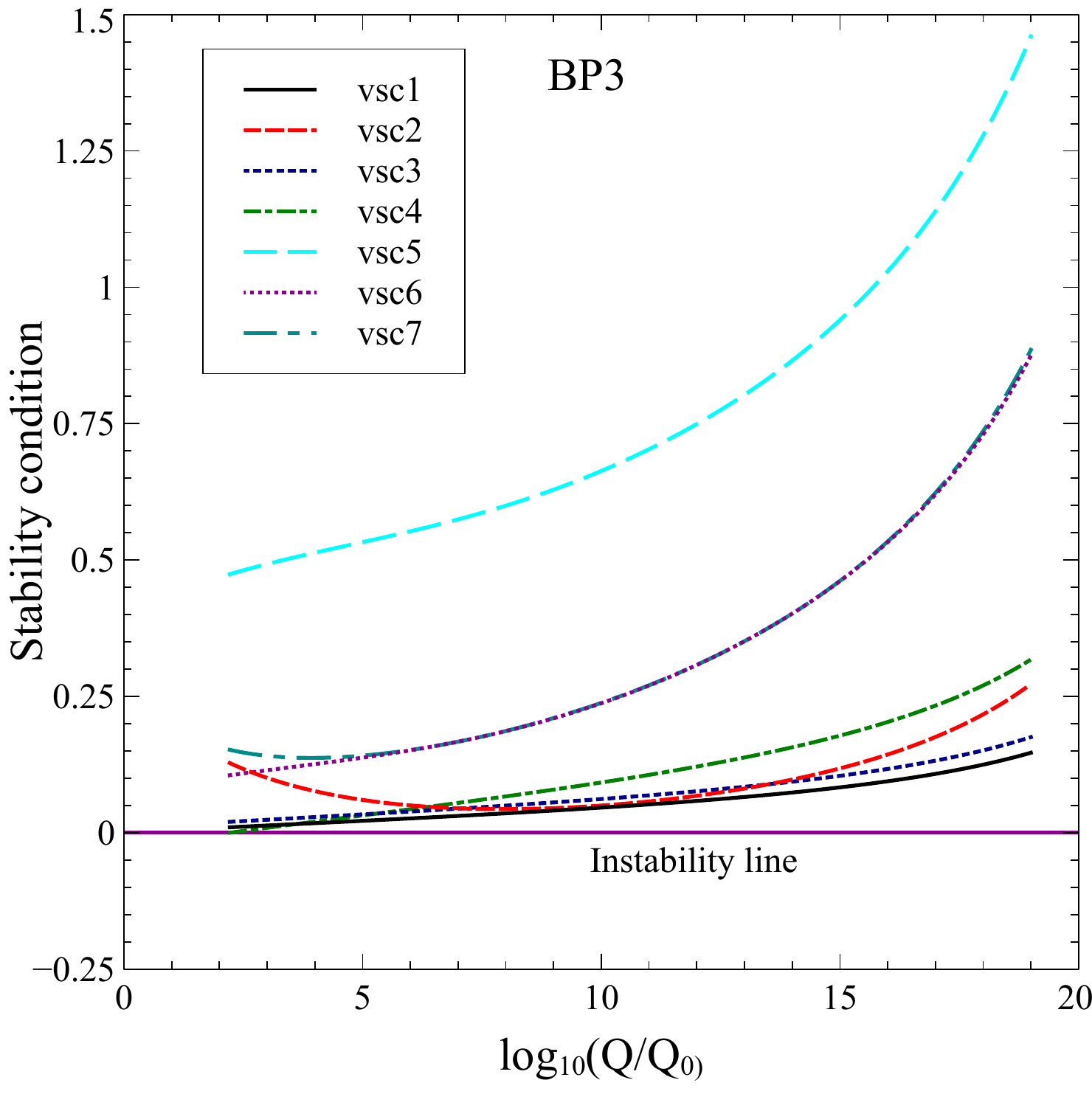}~~~ 
\caption{Evolution of BP3 under RG. Colour coding is explained in the legends and the vacuum instability line is highlighted.}
\label{f:running}
\end{center}
\end{figure}

\subsubsection{\textbf{$m_{H_1} \textgreater ~\rm 370 ~GeV$.}}

In this region,
dark matter relic density tends to diminish due to prohibitively
large annihilation to $VV$ final states. 
Annihilations in this case are the interference of $H_1-H_1-V-V$ four-point coupling
and the $t/u$ channel diagrams with $H^+_1/A_1$ in the propagator. However, a small 
splitting among the masses of the inert scalars induces cancellation between these two classes of diagrams thereby burgeoning relic density to the desired range. Larger is $m_{H_1}$, higher is the 
annihilation to the longitudinal gauge bosons and hence higher becomes $\l_L$. While a similar
phenomenology occurs in case of a single inert doublet, apart from the DM mass $\textless$ 80 GeV region, $\Omega h^2$ is $\sim$ 0.1 again only when
the DM mass $\textgreater$ 500 GeV.
For example, for $m_{H_1}$ = 387.5, $m_{A_1}$ = 390.5, $m_{H^+_1}$ = 389.6, $\l_L$ = 0.056,
the dominant annihilation channels are $H_1 H_1$ $\rightarrow$ $WW$ 12$\%$, $H_2 H_2$ $\rightarrow$ $WW$ 12$\%$, $H_1 H_1$ $\rightarrow$ $ZZ$ 10$\%$, $H_2 H_2$ $\rightarrow$ $ZZ$ 10$\%$, $H^+_1 H^-_1$ $\rightarrow$ $WW$ 6$\%$, $H^+_2 H^-_2$ $\rightarrow$ $WW$ 6$\%$, $H^+_1 H_1$ $\rightarrow$ $\gamma W^+$ 6$\%$, $H^+_2 H_2$ $\rightarrow$ $\gamma W^+$ 6$\%$. For a spectrum $m_{H_1}$ = 904.1, $m_{A_1}$ = 912.1, $m_{H^+_1}$ = 904.3, the requisite $\l_L$ for a correct relic increases to $\sim$ 0.49. One thus requires a small mass splitting and an appropriately adjusted $\l_L$ to generate correct 
relic density. 

To examine high-scale validity of this scenario, model points are generated in the following range.  
\bea
\l_L \in [-4\pi,4\pi] \\
m_{H_1} \in  [300.0 ~\rm GeV, 1000.0~\rm GeV] \\
m_{A_1} \in [ m_{H_1}, m_{H_1} + 100.0 ~\rm GeV]\\
m_{H^+_1} \in [ m_{H_1}, m_{H_1} + 100.0 ~\rm GeV]
\eea

\begin{figure}[h!t]
		\begin{center}
			\subfloat[ \label{sf:relic_mh1}]{%
				\includegraphics[scale=0.45]{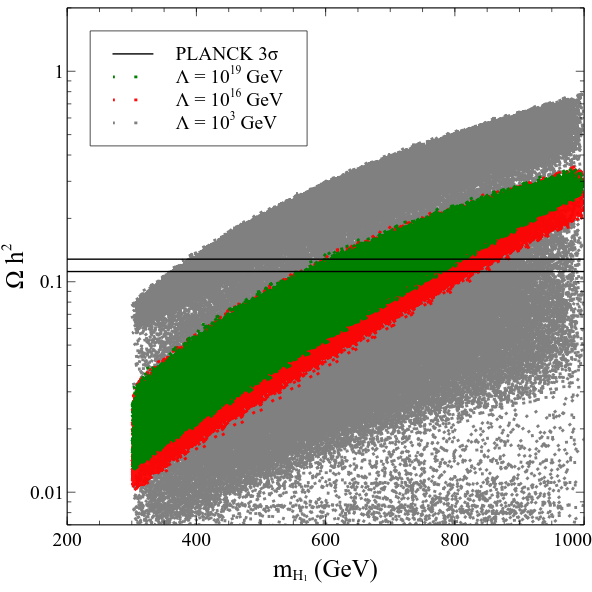}
			}~~~~
			\subfloat[\label{sf:relic_lL}]{%
				\includegraphics[scale=0.45]{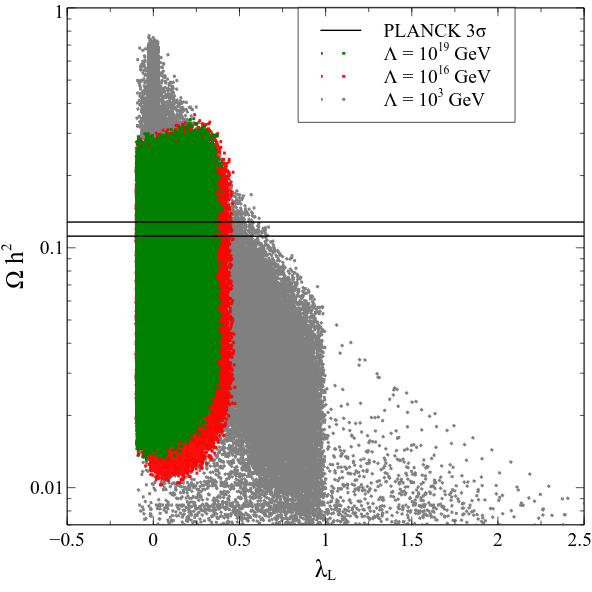}~~~~
			} 			
			\caption{The dark matter relic density versus $m_{H_1}$ (left) and the coupling of $H_1$
pair to the Higgs boson $\l_L$ (right). The grey, green and red points preserve validity up to 1TeV, the GUT scale $10^{16}$ GeV and the Planck scale $10^{19}$ GeV respectively. The horizontal black lines denote the 3$\sigma$ limits of the PLANCK data.
.}
			\label{f:relic}
		\end{center}
	\end{figure}

\begin{figure}[h!t]
		\begin{center}
			\subfloat[ \label{sf:DD_mh1}]{%
				\includegraphics[scale=0.45]{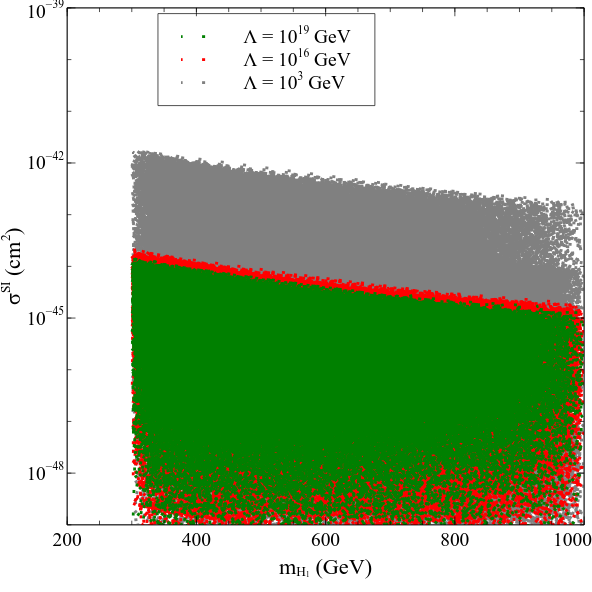}
			}~~~~
			\subfloat[\label{sf:DD_lL}]{%
				\includegraphics[scale=0.45]{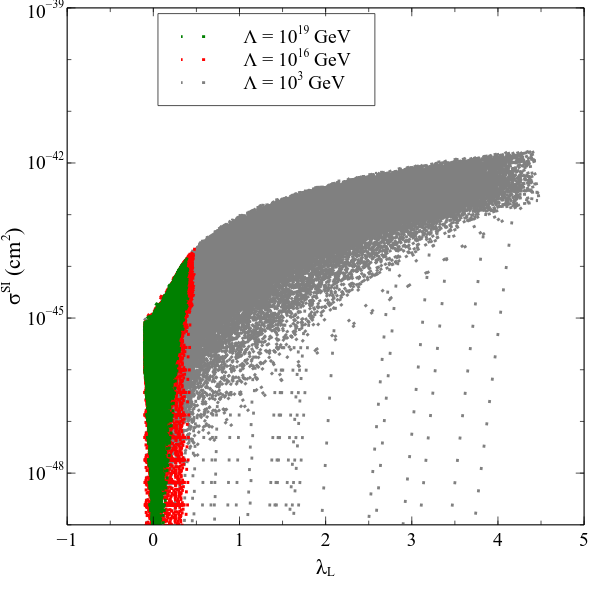}~~~~
			} 			
			\caption{Spin independent WIMP-nucleon scattering cross section vs $m_{H_1}$ (left) and the coupling of $H_1$
pair to the Higgs boson $\l_L$ (right). The grey, green and red points preserve validity up to 1TeV, the GUT scale $10^{16}$ GeV and the Planck scale $10^{19}$ GeV respectively. Note that a large proportion of model points do obey the LUX upper bound while fulfilling stability requirements.}
			\label{f:DD}
		\end{center}
	\end{figure}

We also fix $\l_1$ =  $\l_2$ = $\l_3$ = 0.01 at the initial scale, since
these couplings do not enter into the calculations of relic density and WIMP-nucleon cross sections. This choice is rather judicious, an higher value mostly makes the couplings non-perturbative at high scales.  Fig.\ref{f:relic} displays the variation of relic density corresponding to
model points valid up to three different cut-offs $\L = 10^{3}$ GeV, $10^{16}$ GeV and $10^{19}$ GeV.
Fig.\ref{f:DD} projects spin-independent WIMP-nucleon cross section. 

An inspection of Fig.\ref{f:relic} and Fig.\ref{f:DD} points out that one can render
the $S_3$HDM stable upto GUT and Planck scales with initial conditions consistent with
the observations of relic density and direct detection. We highlight this fact as the 
most important conclusion in this part. This, however happens only if
$m_{H_1} \textgreater$ 550 GeV. This result be understood as follows. The evolution of
$\l_8$ and hence vacuum stability is crucially dictated by the values of $\l_5$, $\l_6$ 
and $\l_7$ at the initial scale. They can be expressed in terms of the masses as,
\bea
\l_5 &=& \l_L + \frac{2}{v^2}(m_{H^+_1}^2 - m_{H_1}^2) \\
\l_6 &=& \frac{1}{v^2}(m_{H_1}^2 + m_{A_1}^2 - 2 m_{H^+_1}^2) \\
\l_7 &=& \frac{1}{2 v^2}(m_{H_1}^2 - m_{A_1}^2)
\eea
We find that for an $H_1$ below 600 GeV, $\l_5$, $\l_6$ and $\l_7$ are not sizable enough
to ensure $\l_8(Q) \textgreater 0$ up to the GUT scale. On the other hand, perturbative
unitarity restricts the mass splitting to $\sim$ 50 GeV which is automatically consistent
with the $T$ parameter constraint. While the stability condition $\l_5 +2\sqrt{\l_8(\l_1+\l_3)} 
~\textgreater ~0$ disfavours large negative values of $\l_5$, tight upper bounds are imposed 
by perturbative unitarity. This translates into $-0.1 \textless \l_L \textless 0.4$
for a model valid up to $M_{Pl}$ (see Fig.\ref{f:relic}).

\begin{figure}[h!t]
		\begin{center}
			\subfloat[ \label{sf:hs_all_lL_mh1}]{%
				\includegraphics[scale=0.45]{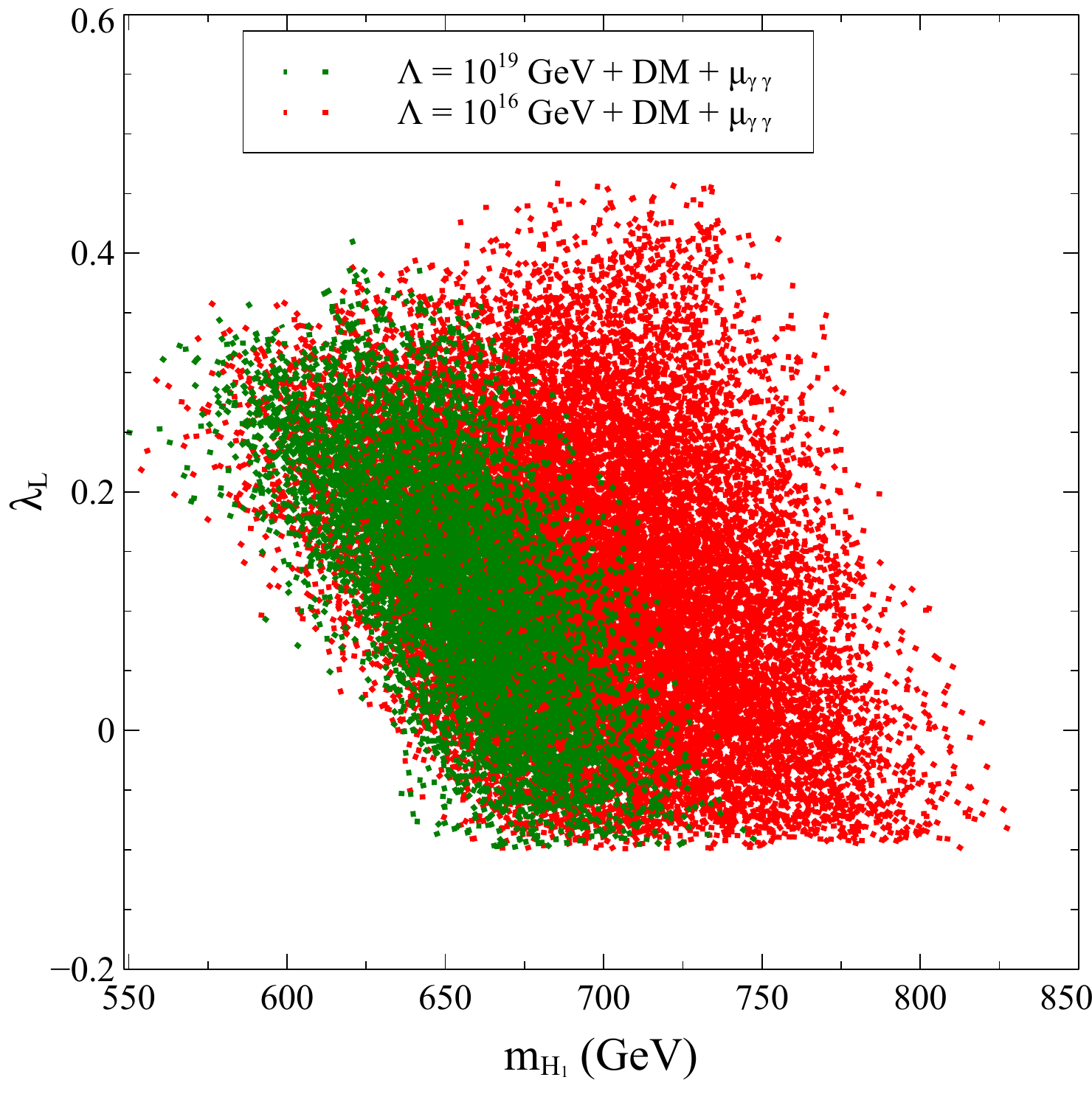}
			}~~~~
			\subfloat[\label{sf:hs_all_ma1mmh1_mh1}]{%
				\includegraphics[scale=0.45]{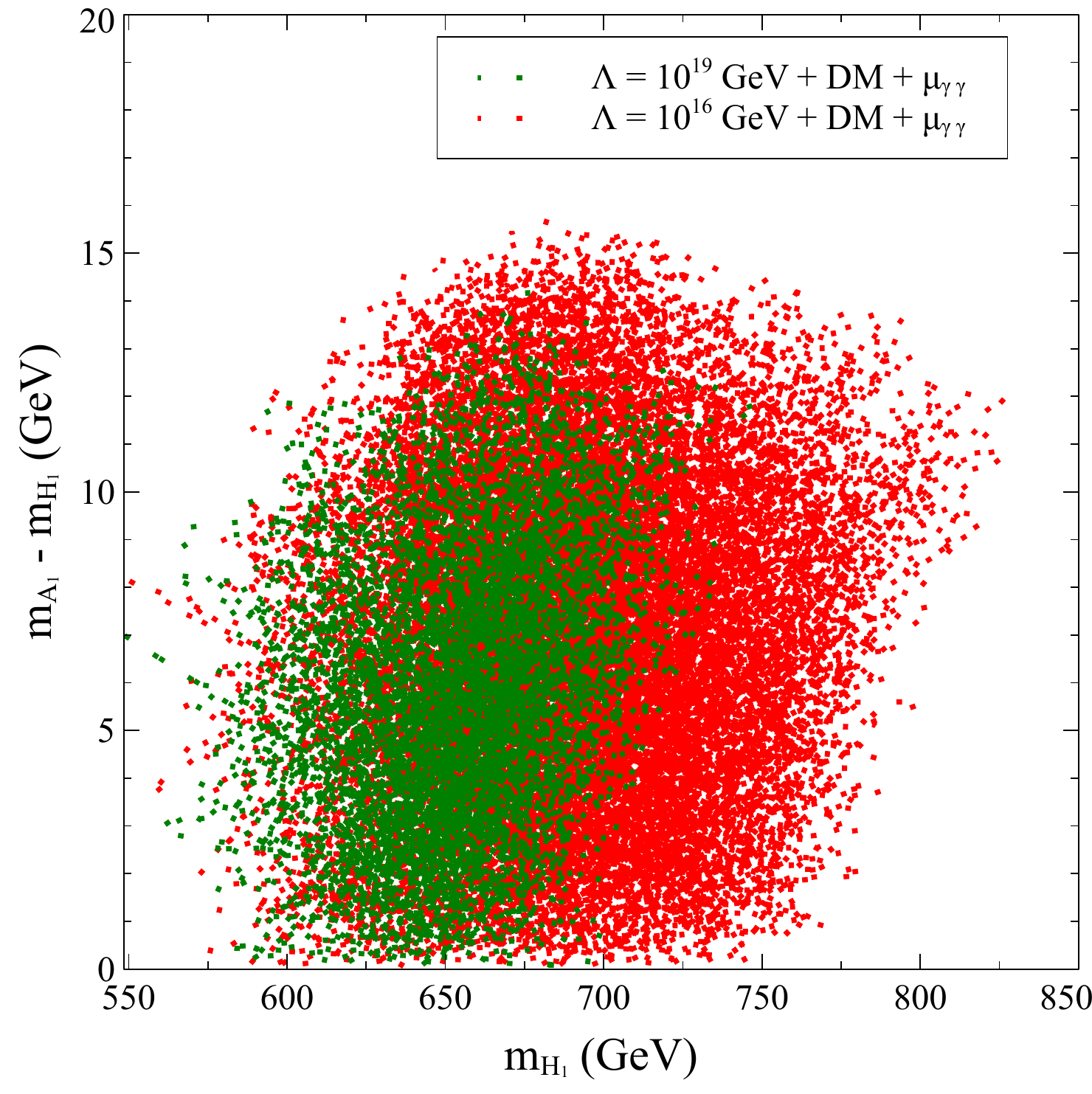}~~~~
			}\\ 
			
			\subfloat[\label{sf:hs_all_mhc1mmh1_mh1}]{%
				\includegraphics[scale=0.45]{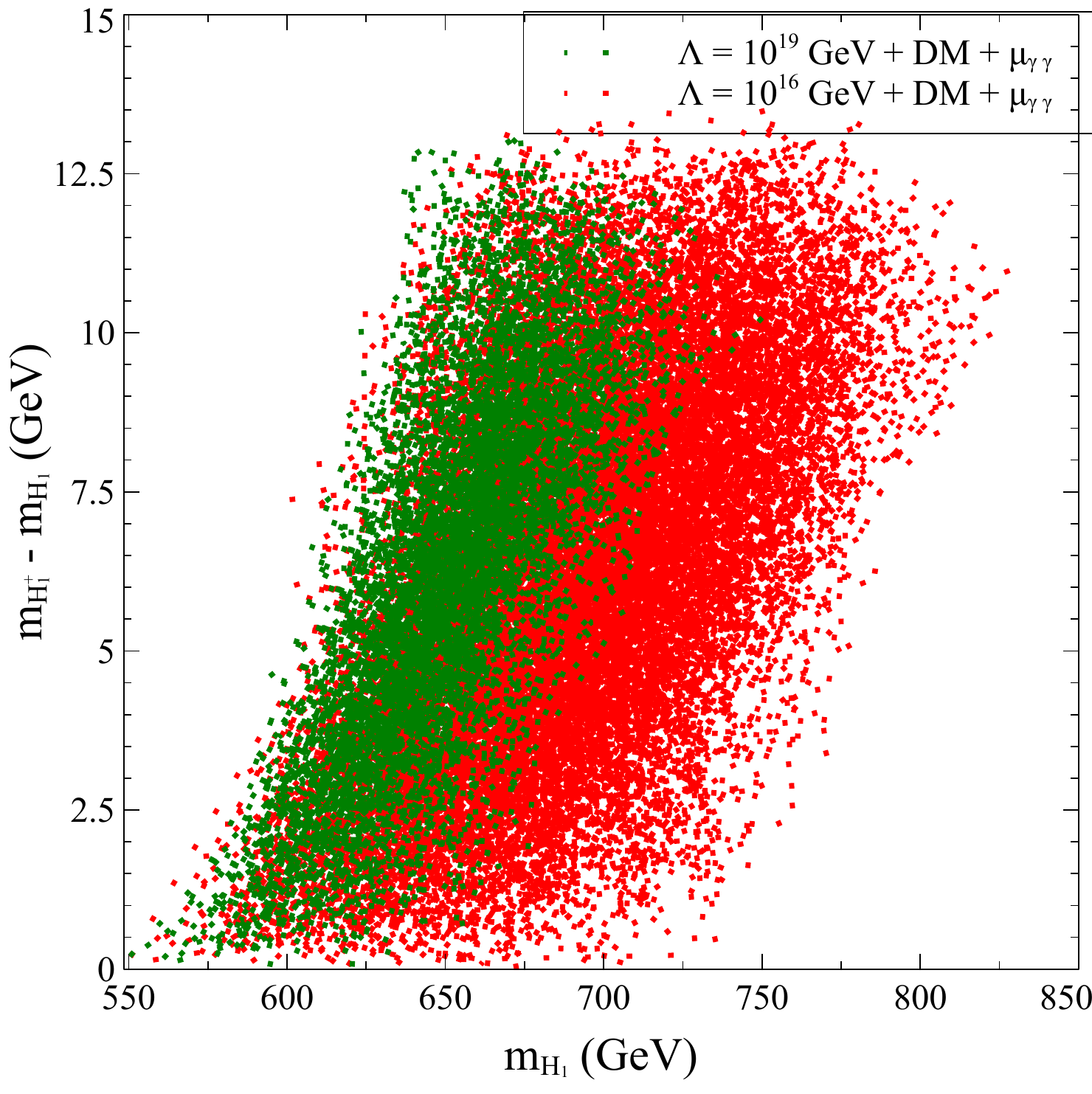}
			}			
			\caption{The viable $S_3$HDM parameter space projected on the $\l_L$ vs $m_{H_1}$ (top left), 
$m_{A_1}-m_{H_1}$ vs $m_{H_1}$ (top right), $m_{H^+_1}-m_{H_1}$ vs $m_{H_1}$ (bottom) planes. "$\L$ + DM + $\mu_{\gamma \gamma}$" in the legends refers to validity up to $\L$ as well as consistency with DM searches and diphoton signal strength. The green and red points correspond to $\L = 10^{16}$ GeV and $\L = 10^{19}$ GeV respectively.} 
			\label{f:hs_all}
		\end{center}
	\end{figure}

For a more comprehensive understanding, the parameter space negotiating all the imposed constraints successfully	 is displayed as correlation plots in Fig.\ref{f:hs_all}. Our demand of $\sigma^{\rm SI} \textless 10^{-46}$ $\rm cm^{2}$ throughout in Fig.\ref{f:hs_all} automatically complies with the LUX results. The DM masses are strongly restricted by the requirements of DM searches, and high-scale validity till a given $\L$. For instance we note $m_{H_1} \in$ [550 GeV,~830 GeV] and [550 GeV,~750 GeV] 
for $\L$ = $M_{\rm GUT}$ and $\L$ = $M_{Pl}$ respectively (see Fig.\ref{f:hs_all}).

\begin{figure}[h!t]
		\begin{center}
			\subfloat[ \label{sf:mugaga_DM}]{%
				\includegraphics[scale=0.45]{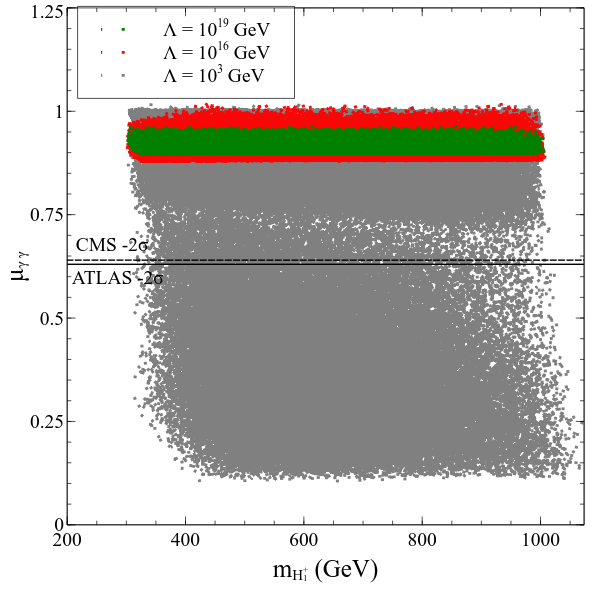}
			}~~~~			
			\caption{Distribution of parameter points valid till $10^3$ GeV (grey), $10^{16}$ (red) GeV and $10^{19}$ (green) GeV in the $\mu_{\gamma\gamma}$ vs $m_{H^+_1}$ plane. The solid and dotted lines denote the 2$\sigma$ limits below the central value given by ATLAS and CMS respectively
.}
			\label{f:mugaga_DM}
		\end{center}
	\end{figure}

A situation, where $\mu_{\gamma\gamma} \textless 1$ (Fig.\ref{f:mugaga_DM}) for most part of the parameter space is attributed to a mostly non-negative $\l_5$ (or a very small negative value).
The reader should note that unlike the previous case, one can in principle have $\mu_{\gamma\gamma} \textgreater 1$ in this case, however subject to stability constraints.
With $\l_1 = \l_3 = 0.01$ at the input scale, $\l_5$ gets bounded from below at $\simeq$ -0.1
by the vacuum stability condition $\l_5 \textgreater -2\sqrt{\l_8(\l_1+\l_3)}$. 
One can get a deeper lower bound, and hence a $\mu_{\gamma\gamma}$ substantially larger than unity for larger values of $\l_1$ and $\l_3$, but in that case 
one does not have a perturabtive theory till $10^{19}$ GeV. Very low values of $\mu_{\gamma\gamma}$ seen in Fig.\ref{f:mugaga_DM} are possible for points valid only up to
the TeV scale, where a positive $\l_5$ as large as $\sim$ 6.5 is allowed without causing a breakdown of perturbativity below 1TeV. Parameter points valid till the GUT and Planck scales rarely correspond a diphoton signal strength less than 0.87. This indeed is within the 2$\sigma$ limit from both the ATLAS and CMS central values. This very observation that validity till very high scales always predicts a depletion in the diphoton rate, but still can be kept within the experimental bounds emerges as an important consequence in this regard. The diphoton rate thus bears fingerprints of an extended
Higgs sector such as the $S_3$HDM, whose tree level couplings could mimic the corresponding SM ones. This calls for its accurate measurements in 13 TeV LHC for instance, or at the other upcoming colliders.

\begin{figure} %[t]
\begin{center}
%\rotatebox{90}{\quad\quad\quad\quad$m_A$ (GeV)}
%
\includegraphics[scale=0.45]{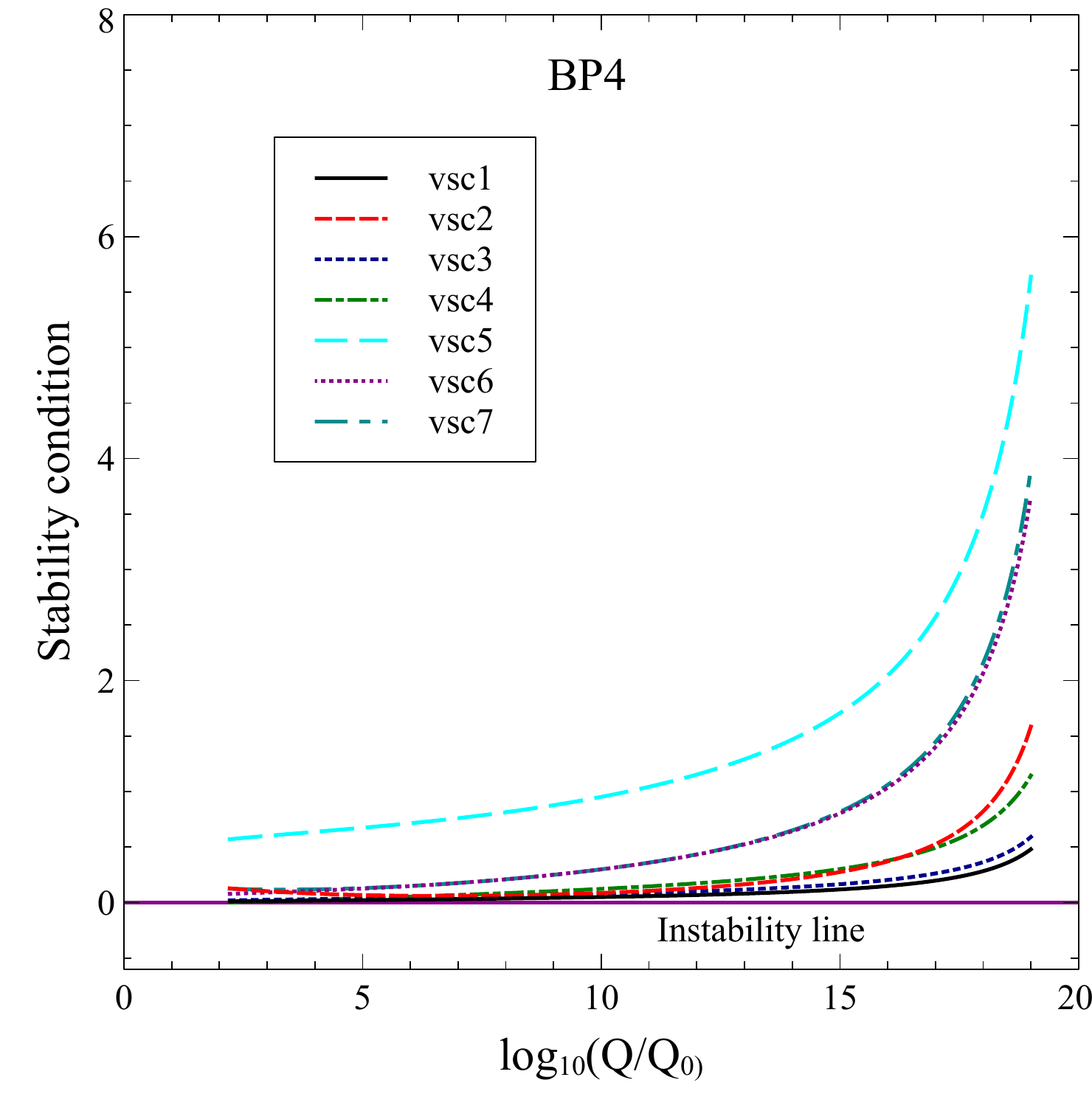}~~~ 
\includegraphics[scale=0.45]{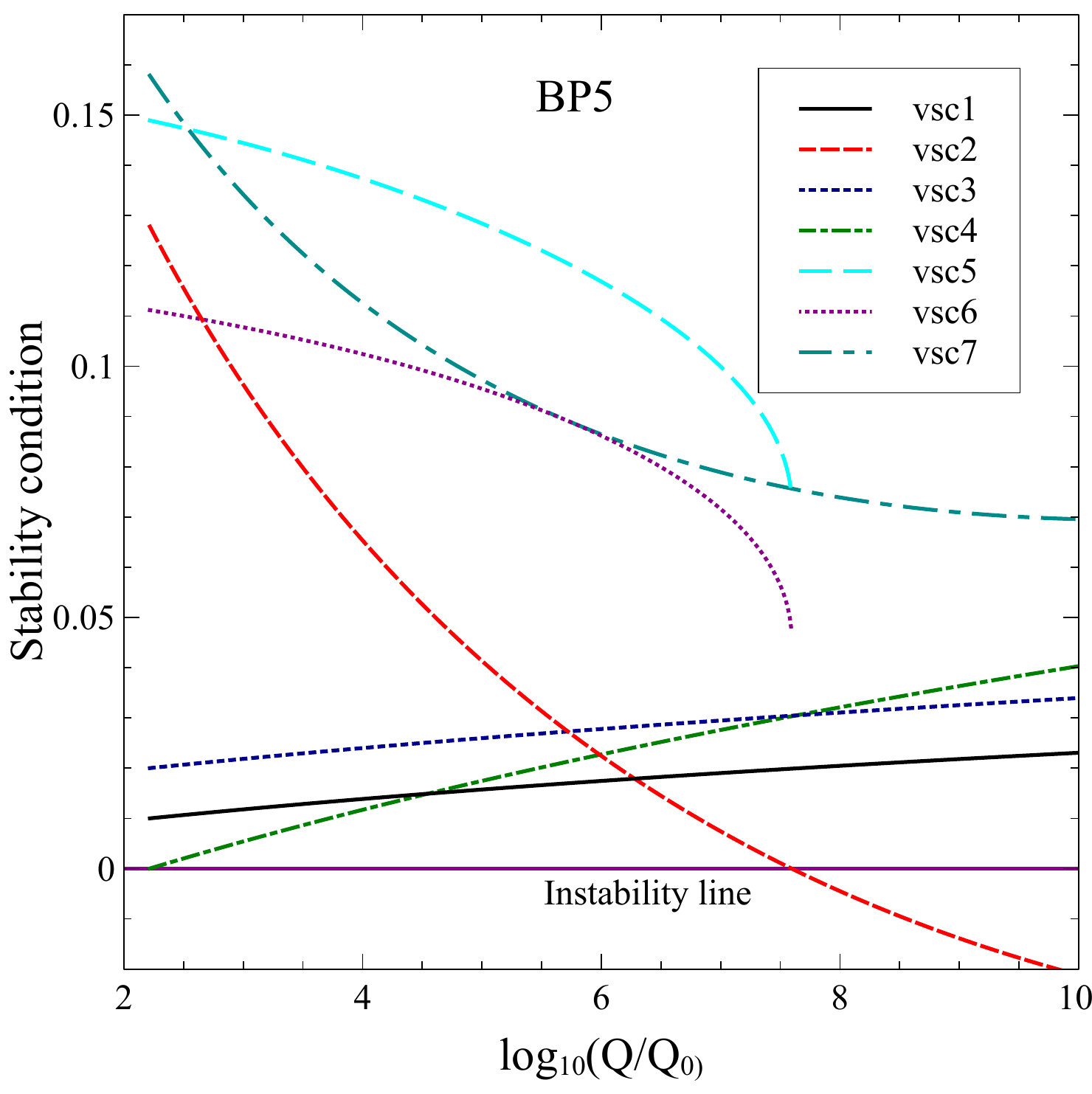}
\\
\includegraphics[scale=0.45]{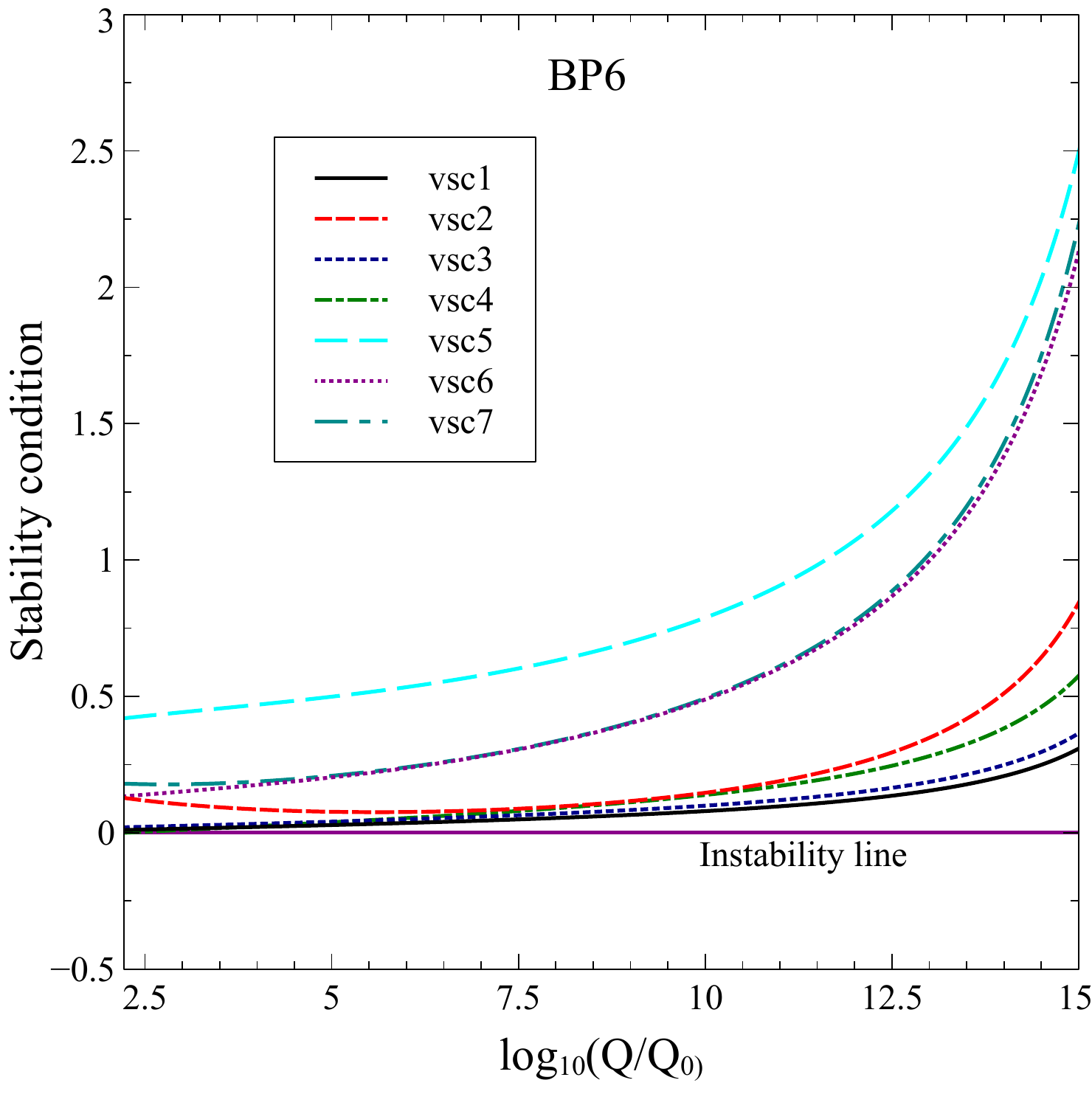}~~~ 
\includegraphics[scale=0.45]{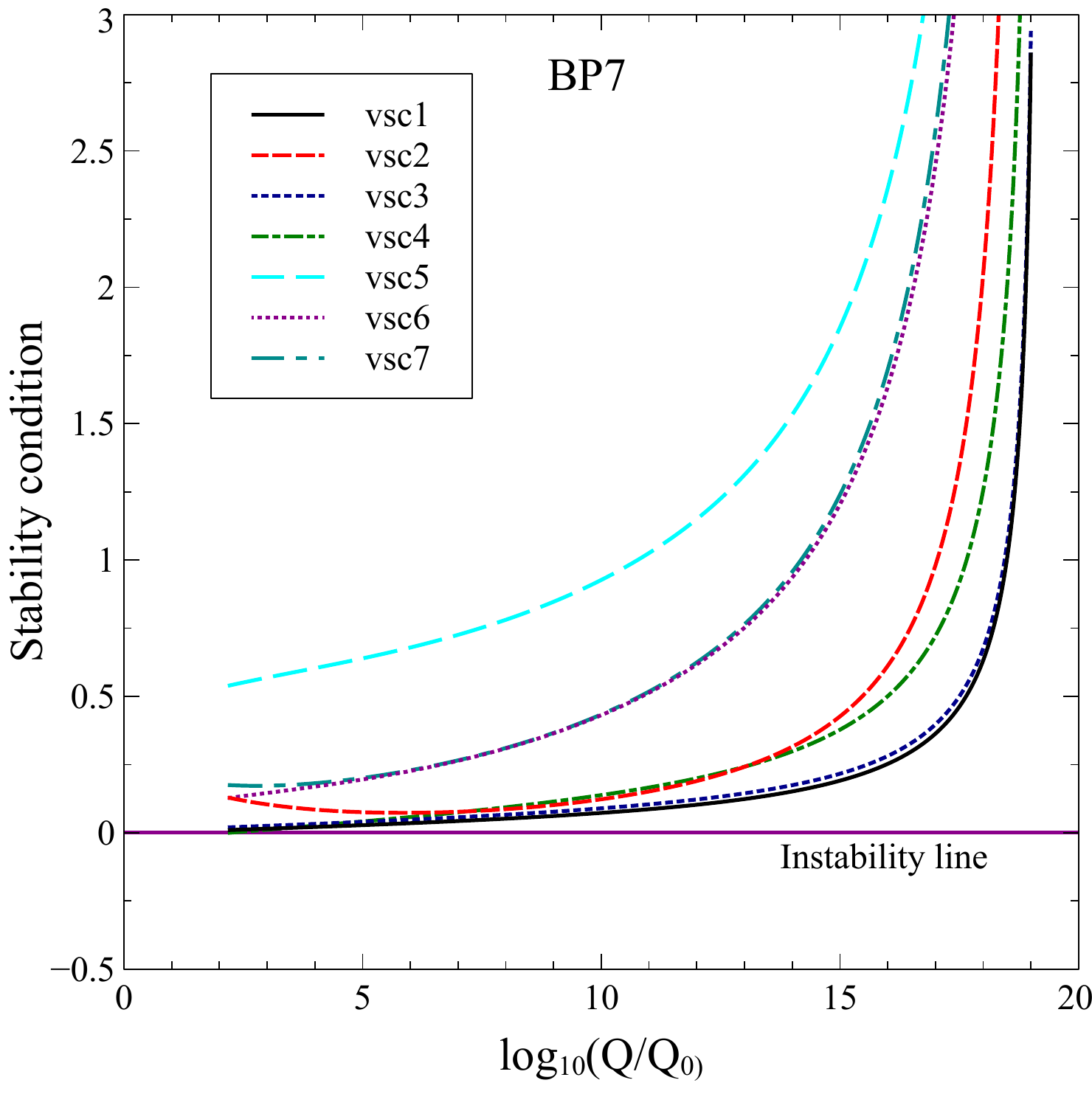}
\caption{RG Evolution of BP4, BP5, BP6 and BP7. Colour coding is explained in the legends and the vacuum instability line is highlighted. Note that vsc5 and vsc6 are not defined whenever $\l_8$ $\textless$ 0.}
\label{f:BP_DM}
\end{center}
\end{figure}

To sum up, DM phenomenology plays a vital role in deciding the fate of this scenario at high scales. The interplay of various effects involved is captured through the benchmarks in Table~\ref{BP4to7}. The 
RG running of these benchmarks is shown in Fig.\ref{f:BP_DM} The first benchmark BP4 can possibly describe physics nearly up to the GUT scale, beyond which perturbativity breaks down. However, BP4 predicts a relic density below the observed limit.
This is attributed to the relatively large mass splittings amongst the $S_3$ scalars, which generate
such sizable $\l_5$, $\l_6$ and $\l_7$ at the initial scale that can ensure $\l_8$(Q) $\textgreater$ 0 throughout. However we pay the price of a diminished co-annihilation, and thereby a relic density below the desired range. A fall out of this relatively large $\l_5$ in this case is a suppressed $\mu_{\gamma \gamma}$. BP5 highlights the fact that correct relic density and direct detection rates are achievable in this model for a DM around 390 GeV, a feature not observed in the model with a single inert doublet. The maximal mass difference in such a case is restricted to $\sim$ 13 GeV. However, BP5 does not keep the EW vacuum stable beyond $10^8$ GeV.  
 
\begin{table}[h]
\centering
\begin{tabular}{|c c c c c c c c|}
\hline
Benchmark  & $m_{H_1}$(GeV) & $m_{A_1}$(GeV) & $m_{H^+_1}$(GeV)  & $\l_L$  & $\Omega h^2$ & $\sigma_{SI} (cm^2)$   & $\L$(GeV)\\ \hline \hline

BP4 & 479.200 & 480.475 & 494.525 & -0.0236 & 0.0635 & 2.13 $\times$ $10^{-47}$ &  $10^{19}$\\ \hline
BP5 & 390.000 & 391.000 & 392.000 & 0.0050 & 0.1200 & 1.44 $\times$ $10^{-48}$ & $10^8$\\ \hline
BP6 & 707.400 & 720.000 & 713.500	& 0.032	 & 0.1214 & 1.80 $\times$ $10^{-47}$ & Just below $10^{16}$ \\ \hline
BP7 & 718.600 & 727.450		& 727.225		& 0.0268 & 0.1263 & 1.22 $\times$ $10^{-47}$ & $10^{19}$ \\ \hline

\end{tabular}
\caption{Benchmark points chosen to illustrate the behaviour under RGE. $\L$ denotes the maximum
extrapolation scale up to which vacuum stability and perturbativity are ensured.}
\label{BP4to7}
\end{table}

BP6 and BP7 are conservative choices which predict relic density and direct-detection rates in the
correct ballpark, and also extrapolate the model to the GUT and Planck scales respectively. We note here that in BP4, BP6 and BP7, vsc1, vsc3, vsc4, vsc5, vsc6, vsc7 rise with $Q$, whereas in BP5, vsc5 and vsc6 go down. This
observation has its root in the structure of the $S_3$HDM beta functions (see Appendix.\ref{ss:RGE}), which mostly guarantee 
vsc1, vsc3, vsc4, vsc5, vsc6, vsc7 $\textgreater$ 0 throughout the evolution once they start with positive values at the EW scale.
We remark here that BP6 and BP7 correspond to $\mu_{\gamma \gamma}$ = 0.935 and 0.911 respectively,which are within the 2$\sigma$ limit from the central value.

In the same connection, we have found that an $m_{H_1} \textgreater$ 1 TeV can
render the EW vacuum stable at least up to the SM instability scale. However that pushes $\mu_{11}$ to yet higher values, thereby
introducing a so-called \emph{intermediate} scale into the picture. It is then implied that the $S_3$
scalars are practically decoupled below $\mu_{11}$, and that it would be more appropriate to solve the
RG equations in a piecewise fashion, i.e., evolve from the EW scale to $\mu_{11}$ using the SM beta
functions only, and then invoke $S_3$HDM above the $\mu_{11}$ threshold. However we mostly encounter $\mu_{11}$ $\lesssim$ 600 GeV for $S_3$ masses $\textless$ 1 TeV. We have checked that for such a  $\mu_{11}$, a piecewise evolution practically gives the same numerical results.

\section{Conclusions and future work}\label{Conclusions} 

3HDMs offer a rich scalar spectrum and can give rise to prominent signatures at the colliders\cite{Bhattacharyya:2012ze,Bhattacharyya:2010hp}.
In this paper, we have tried to investigate an $S_3$-symmetric Higgs sector in the light of various theoretical as well as experimental constraints. Robust regimes of the model parameter space were surveyed using the latest data on the 125 GeV Higgs and oblique parameters. The high-scale
behaviour was probed by evolving the model couplings under the RGEs, and this study appears to 
be the first attempt in that direction in context of 3HDMs. A unitary and 
perturbative theory,
along with a stable EW vacuum was ensured at each step of evolution. We have illustrated our 
findings in context of two specific alignment of the doublet vevs. The salient features of the 
numerical results that emerge are highlighted below. 

\begin{itemize}

\item In the first case, non-zero vevs are assigned to all three of the doublets while 
maintaining $v_1=\surd3 v_2$.
It is found that this scenario is not stable beyond $10^7$ GeV, an effect brought about
by an interplay of perturbative unitarity and vacuum stability.
Stringent upper bounds are placed on the scalar masses and tan$\beta$ in this case. 
In particular we note tan$\beta$ $\textless$ 1.3 and the $S_3$ scalar masses lie below 270 GeV
for $\L$ = $10^6$ GeV, the maximum phenomenologically accepted scale. 

\item The second case is a scenario with two inert doublets. 
There lies an identifiable region in the parameter space in this 
case that extends validity of the model till the Planck scale. 
Moreover, this parameter space is robust enough to accomodate a successful
canditate for dark matter. High-scale stability in this case manifests
itself by placing upper bounds on the coupling of the DM to the 
125 GeV Higgs, the DM mass, as well on the mass splitting amongst the inert scalars.
The bounds get sharper when both the DM as well as high-scale stability constraints
are imposed simultaneously. In a word, a connection emerges between DM phenomenology 
at the \emph{low} scale and a good UV behaviour at \emph{high} scales. This finding
is qualitatively similar to the model with a single inert doublet\cite{Goudelis:2013uca}.
However, the addition of the extra inert doublet narrows down the gap between the 
\emph{low} and \emph{high} DM mass regions, with respect to what is observed in the single 
inert doublet case.

\item Scenario B predicts a \emph{decrement} in the diphoton decay width with respect to the SM value, so does Scenario A for an exact alignment. This particular
feature of Scenario B is not seen by considering tree-level stability constraints alone and is an explicit
consequence of renormalisation group evolution. The numerical predictions however 
can be made to lie within the current experimental limits without running into conflict
with high-scale stability.  
\end{itemize}

Altogether then, we conclude that the inert scenario fares much better 
than the non-inert one in terms of high-scale validity and signal strength measurements. 
It is thus safe to comment that the $S_3$HDM can certainly alleviate the vacuum stability problem, 
however not for all permissible vev structures. Several extensions
of the present study are possible. One could analyze a more general $S_3$-symmetric 
Yukawa texture in a similar context, admittedly though such texture would give rise to
Flavor-Changing Neutral Currents (FCNC)\cite{Mondragon:2007af} at the tree level. It was
shown in~\cite{PAKVASA197861} that raising the $S_3$ masses to $\sim$10 TeV suppresses the 
possible FCNCs.
The requirement of such heavy scalars necessiates the inclusion of $S_3$ violating quadratic terms\cite{Kubo:2004ps}.
Another motivation of a broken $S_3$ symmetry is in the context of DM. In scenario B for instance,
it will lead to a non-degenrate spectrum and hence a modified DM phenomenology, at least at the
quantitative level. Indirect detection signatures of such a DM scenario could be of special importance in light of latest data. Adding further to it, the large number of bosonic degrees of freedom offered by the $S_3$HDM  could favour a \emph{strong} first order electroweak phase transition, thereby making way for baryogenesis, something already looked at for a more generic
3HDM with two inert doublets.~\cite{Ahriche:2015mea}.

\section{Acknowledgements}

I thank Aritra Gupta for useful discussions, Arijit Dutta for a computational aid, and Biswarup Mukhopadhyaya for his insightful comments on the manuscript. I also acknowledge the funding available from the Department of Atomic Energy, Government of India, for the Regional Centre for Accelerator based Particle Physics (RECAPP), Harish-Chandra Research Institute.

\section*{Appendix.}\label{Appendix}
\appendix
\label{s:appen}
\section{Renormalisation group (RG) equations}
%\numberwithin{equation}{section}
\label{ss:RGE}
We list the one-loop RG equations for the model couplings used throughout the analysis.
For the gauge couplings, they are given by~\cite{Branco:2011iw},
\besub
\bea
16\pi^2 \frac{dg_s}{dt} &=& - 7 g_s^3,
\\
16\pi^2 \frac{dg}{dt} &=& - \frac{17}{6} g^3,
\\
16\pi^2 \frac{dg^{\prime}}{dt} &=& \frac{43}{6} {g^\prime}^3.
\eea
\eesub

The quartic couplings evolve according to,
\besub
\bea
\label{e:dl1}
16\pi^2 \beta_{\l_1}  &=&
32\l_1^2 + 8\l_2^2 + 16\l_3^2 + 4\l_4^2 + 2\l_5^2 + \frac{1}{2}\l_6^2 - 8\l_1\l_2 + 16\l_1\l_3\nonumber \\
 & &
+ 2\l_5\l_6 + 2\l_7^2 + \frac{3}{8}(g^{\prime 4} + 3g^{4})- \lambda_1 (9g^2 + 3g^{\prime 2} )\
\label{eq:dl2} \\
16\pi^2 \beta_{\l_2}  &=&
24\l_1\l_2 - 24\l_2^2 - 16\l_2\l_3 - \frac{1}{2}\l_6^2 + 2\l_7^2 - \frac{3}{4}g^{\prime 2}g^{ 2}\nonumber \\
 & &
 - \lambda_2 (9g^2 + 3g^{\prime 2} )\
\label{eq:dl3} \\
16\pi^2 \beta_{\l_3}  &=&
16\l_3^2 + 8\l_4^2 + 24\l_1\l_3 + 8\l_2\l_3 + 8\l_4^2 + \frac{1}{2}\l_6^2 + 2\l_7^2 + \frac{3}{4}g^{\prime 2}g^{ 2}\nonumber \\
 & &
 - \lambda_3 (9g^2 + 3g^{\prime 2} )\
\label{eq:dl4} \\
16\pi^2 \beta_{\l_4}  &=&
\lambda_4( 12 \lambda_1 + 4 \lambda_2 + 24 \lambda_3 + 6 \lambda_5 + 8 \lambda_6 +20 \lambda_7)  \nonumber \\
 & &
- \lambda_4 (9g^2 + 3g^{\prime 2}- 3 y_t^2 )\,, \label{eq:dl5} \\
16\pi^2 \beta_{\l_5}  &=&
4\l_5^2 + 2\l_6^2 + 8\l_4^2 + 8\l_7^2 + 20 \l_1\l_5 - 4\l_2\l_5 + 8\l_3\l_5 + 8\l_1\l_6\nonumber \\
 & &
+ 12 \l_5\l_8 + 4\l_6\l_8 + \frac{3}{4}(g^{\prime 4} - 2g^{\prime 2}g^{ 2}+3g^4)- \lambda_5 (9g^2 + 3g^{\prime 2}- 6 y_t^2 )\,, \\
16\pi^2 \beta_{\l_6}  &=&
20\l_4^2 + 4\l_6^2 +32\l_7^2 + 4\l_1\l_6 - 4\l_2\l_6 + 8\l_3\l_6 + 8\l_5\l_6 + 4\l_8\l_6 + 3g^{\prime 2}g^{2} \nonumber \\
& &
- \lambda_6 (9g^2 + 3g^{\prime 2}- 6 y_t^2 )\,, \\
16\pi^2 \beta_{\l_7}  &=&
4\l_7(\l_1 + \l_2 + 2\l_3 + 2\l_5 + 3\l_6 + \l_8) + 10\l_4^2 \nonumber \\
 & &
-\ \lambda_7 (9g^2 + 3g^{\prime 2}- 6 y_t^2 )\,, 
\label{eq:dl6} \\
16\pi^2 \beta_{\l_8}  &=&
4\l_5^2 + 4\l_5\l_6 + 2\l_6^2 + 8\l_7^2 + 24\l_8^2 + \frac{3}{8}(g^{\prime 4} + 2g^{\prime 2}g^{ 2}+3g^4) \nonumber \\
 & &
-\ \lambda_8 (9g^2 + 3g^{\prime 2}- 12 y_t^2 ) - 6 y_t^4\,,
\label{eq:dl7} 
\eea
\eesub
Neglecting the effect of other quarks, the t-quark Yukawa coupling has the beta
function,
\bea
 16\pi^2 \beta_{y_t} &=& y_{t}\left(-8g_s^2 - \frac94 g^2 - \frac{17}{12} g^{\prime 2}+
 \frac92 y_{t}^2\right)\,
\eea

\section{Oblique parameters}
\label{ss:Oblq}

The expressions for the oblique parameters in the $S_3$HDM are given. A shorthand notation
sin$(\beta - \alpha)$ = $s_{\beta - \alpha}$, cos$(\beta - \alpha)$ = $c_{\beta - \alpha}$
is adopted,
\besub
\bea
\Delta S &=& (2 s_{W}^2-1)^2 G(m_{H^+_1}^2,m_{H^+_1}^2,m_{Z}^2)+(2 s_{W}^2-1)^2 G(m_{H^+_2}^2,m_{H^+_2}^2,m_{Z}^2)+G(m_{H_2}^2,m_{A_1}^2,m_{Z}^2) \nonumber \\
 & &
+c_{\beta - \alpha}^2 G(m_{h}^2,m_{A_2}^2,m_{Z}^2)+s_{\beta - \alpha}^2 G(m_{H_1}^2,m_{A_2}^2,m_{Z}^2)
+c_{\beta - \alpha}^2 G(m_{H_1}^2,m_{H_1}^2,m_{Z}^2)
\nonumber \\
 & &
-s_{\beta - \alpha}^2 G(m_{h}^2,m_{h}^2,m_{Z}^2)
-2{ln(m_{H^+_1}^2)}-2 ln(m_{H^+_2}^2)+ln(m_{H_2}^2)+ln(m_{H_1}^2)+ln(m_{A_1}^2)\nonumber \\
 & &
 + ln(m_{A_2}^2) \\
 \Delta T &=& F(m_{H^+_1}^2,m_{H_2}^2)+F(m_{H^+_1}^2,m_{A_1}^2)+c_{\beta - \alpha}^2 F(m_{H^+_2}^2,m_{h}^2)+s_{\beta - \alpha}^2 F(m_{H^+_2}^2,m_{H_1}^2)-F(m_{H_2}^2,m_{A_1}^2)\nonumber \\
 & &
-c_{\beta - \alpha}^2 F(m_{h}^2,m_{A_2}^2)-s_{\beta - \alpha}^2 F(m_{H_1}^2,m_{A_2}^2)+3c_{\beta - \alpha}^2(F(m_{Z}^2,m_{H_1}^2)-F(m_{W}^2,m_{H_1}^2))\nonumber \\
 & &
-3c_{\beta - \alpha}^2(F(m_{Z}^2,m_{h}^2)-F(m_{W}^2,m_{h}^2)) 
\\
\Delta U &=& \frac{1}{24\pi}[G(m_{H^+_1}^2,m_{H_2}^2,m_{W}^2) + G(m_{H^+_1}^2,m_{A_1}^2,m_{W}^2) + c_{\beta - \alpha}^2 G(m_{H^+_2}^2,m_{h}^2,m_{W}^2)\nonumber \\
 & &
 + s_{\beta - \alpha}^2 G(m_{H^+_2}^2,m_{H_1}^2,m_{W}^2) + G(m_{H^+_2}^2,m_{A_2}^2,m_{W}^2) +  c_{\beta - \alpha}^2\hat G(m_{H_1}^2,m_{W}^2) - \hat G(m_{H_1}^2,m_{Z}^2)\nonumber \\
& &
 - c_{\beta - \alpha}^2\hat G(m_{h}^2,m_{W}^2) - \hat G(m_{h}^2,m_{Z}^2) - G(m_{H_2}^2,m_{A_1}^2,m_{Z}^2) - c_{\beta - \alpha}^2 G(m_{h}^2,m_{A_2}^2,m_{Z}^2)\nonumber \\
& & - s_{\beta - \alpha}^2 G(m_{H_1}^2,m_{A_2}^2,m_{Z}^2) - (2 s_{W}^2-1)^2 G(m_{H^+_1}^2,m_{H^+_1}^2,m_{Z}^2)\nonumber \\
& &
 - (2 s_{W}^2-1)^2 G(m_{H^+_2}^2,m_{H^+_2}^2,m_{Z}^2)]
\eea
\eesub

where, 

\be
F \left( m_{1}^2, m_{2}^2 \right) \equiv \left\{ \begin{array}{lcl}
\displaystyle{
\frac{m_{1}^2 + m_{2}^2}{2} - \frac{m_{1}^2 m_{2}^2}{m_{1}^2 - m_{2}^2}\, \ln{\frac{m_{1}^2}{m_{2}^2}}
}
& ; & m_{1}^2 \neq m_{2}^2,
\\*[3mm]
0 & ; & m_{1}^2 = m_{2}^2.
\end{array} \right.
\label{funcF}
\ee

\bea
G \left( m_{1}^2, m_{2}^2, q^2 \right) &\equiv&
- \frac{16}{3}
+ \frac{5 \left( m_{1}^2 + m_{2}^2 \right)}{q^2}
- \frac{2 \left( m_{1}^2 - m_{2}^2 \right)^2}{(q^2)^2}
\no & &
+ \frac{3}{q^2}
\left[ \frac{m_{1}^4 + m_{2}^4}{m_{1}^2 - m_{2}^2}
- \frac{m_{1}^4 - m_{2}^4}{q^2}
+ \frac{\left( m_{1}^2 - m_{1}^2 \right)^3}{3 q^4} \right]
\ln{\frac{m_{1}^2}{m_{2}^2}}
+ \frac{r}{(q^2)^3}\, f \left( t, r \right)~~~~~~~~~
\label{Gbar}
\eea

\be
\tilde G \left( m_{1}^2, m_{2}^2, q^2 \right) \equiv
- 2 + \left( \frac{m_{1}^2 - m_{2}^2}{q^2} - \frac{m_{1}^2 + m_{2}^2}{m_{1}^2 - m_{2}^2} \right) \ln{\frac{m_{1}^2}{m_{2}^2}}
+ \frac{f \left( t, r \right)}{q^2}.
\ee

\be
\hat G \left( m^2, q^2 \right) \equiv G \left( m^2, m^2, q^2 \right)+12~\tilde G \left( m^2, m^2, q^2 \right)
\ee

\be
\label{tr}
t \equiv m_{1}^2 + m_{2}^2 - q^2
\quad \mbox{and} \quad
r \equiv (q^2)^2 - 2 q^2 \left( m_{1}^2 + m_{2}^2 \right) + \left( m_{1}^2 - m_{2}^2 \right)^2
\ee

\be\label{f}
f \left( t, r \right) \equiv \left\{ \begin{array}{lcl}
{\displaystyle
\sqrt{r}\, \ln{\left| \frac{t - \sqrt{r}}{t + \sqrt{r}} \right|}
} & ; & r > 0,
\\*[3mm]
0 & ; & r = 0,
\\*[2mm]
{\displaystyle
2\, \sqrt{-r}\, \tan^{-1}{\frac{\sqrt{-r}}{t}}
} & ; & r < 0.
\end{array} \right.
\ee

These are standard functions arising in a one-loop calculation.

\section{$h\rightarrow\gamma\gamma$ decay width}\label{h2gg}

The partial decay width of the SM-like Higgs to a pair of photons in this case has the expression\cite{Djouadi:2005gi},
\begin{eqnarray}
 \Gamma (h\to \gamma\gamma) = \frac{\alpha^2g^2}{2^{10}\pi^3}
 \frac{m_h^3}{M_W^2} \Big|sin(\b-\a) F_W + \Big(-\frac{sin\a}{cos\b}\Big)\frac{4}{3}F_t  + \sum_{i=1}^{2}\kappa_{i} F_{i+} \Big|^2
 \,, 
\end{eqnarray}

The functions ${F}_W$, ${F}_t$ and ${F}_{i+}$ capture the effects of a W-boson, a t-quark
and a charged scalar running in the loop and shall be defined as,
\begin{subequations}
\begin{eqnarray}
 F_W &=& 2+3\tau_W+3\tau_W(2-\tau_W)f(\tau_W) \,,  \\
 F_t &=& -2\tau_t \big[1+(1-\tau_t)f(\tau_t)\big] \,,  \\
 F_{i+} &=& -\tau_{i+} \big[ 1-\tau_{i+}f(\tau_{i+}) \big]\,.
\end{eqnarray}
\end{subequations}
\begin{eqnarray}
f(\tau) &=&
\left[\sin^{-1}\left(\sqrt{1/\tau}\right)\right]^2 \,.\\ \rm with,~ 
\tau &=& \frac{4 m^2_{a}}{m^2_h}
\label{f}
\end{eqnarray}

Here, $a$ = $t$, $W$ and $H^{+}_i$.

For Scenario A:
\begin{subequations}
\begin{eqnarray}
\kappa_1 &=& -\frac{1}{6 v}(2  \cos\a \ \textrm{cosec}\b(-6 m^2_{H_1} + 3 m^2_{H_2} - 3 m^2_{h} + m^2_{H_2} + 3 m^2_{H^+_2}  \cos2\b)+ \nonumber \\
& &
(6 m^2_{H^+_2} + 2 m^2_{H_2} + m^2_{H^+_2} \rm cos2\b)\rm sec\b~\rm sin\a)\,,  \\
\kappa_2 &=& \frac{1}{9 v}((9(-2 m^2_{H^+_2} + m^2_h)\cos\b - 9 m^2_{h} \textrm{sec}\b + m^2_{H_2} \rm sec^3\b)\sin\a+ \nonumber \\
& &
((9 m^2_h + m^2_{H_2})\textrm{cosec}\b + 18 m^2_{H^+_2}\sin\b - 9 m^2_{h}\sin\b +m^2_{H_2}\textrm{sec}\b \tan\b )\cos\a)
\end{eqnarray}
\end{subequations}

For Scenario B:
\begin{subequations}
\begin{eqnarray}
\kappa_1 &=& \kappa_2 = -\frac{\l_5}{2}  \,.
\end{eqnarray}
\end{subequations}

%%%%%%%%%%%%%%%%%   References %%%%%%%%%%%%%%%%%%%%%%%%%%%%%%%%%%%%
\bibliographystyle{JHEP}
\bibliography{ref.bib}        

\providecommand{\href}[2]{#2}\begingroup\raggedright\begin{thebibliography}{10}

\bibitem{Aad:2012tfa}
{\bf ATLAS} Collaboration, G.~Aad et~al., {\it {Observation of a new particle
  in the search for the Standard Model Higgs boson with the ATLAS detector at
  the LHC}},  {\em Phys.Lett.} {\bf B716} (2012) 1--29,
  [\href{http://xxx.lanl.gov/abs/1207.7214}{{\tt arXiv:1207.7214}}].

\bibitem{Chatrchyan:2012ufa}
{\bf CMS} Collaboration, S.~Chatrchyan et~al., {\it {Observation of a new boson
  at a mass of 125 GeV with the CMS experiment at the LHC}},  {\em Phys.Lett.}
  {\bf B716} (2012) 30--61, [\href{http://xxx.lanl.gov/abs/1207.7235}{{\tt
  arXiv:1207.7235}}].

\bibitem{Freitas:2012kw}
A.~Freitas and P.~Schwaller, {\it {Higgs CP Properties From Early LHC Data}},
  {\em Phys. Rev.} {\bf D87} (2013), no.~5 055014,
  [\href{http://xxx.lanl.gov/abs/1211.1980}{{\tt arXiv:1211.1980}}].

\bibitem{Djouadi:2013yb}
A.~Djouadi, R.~M. Godbole, B.~Mellado, and K.~Mohan, {\it {Probing the
  spin-parity of the Higgs boson via jet kinematics in vector boson fusion}},
  {\em Phys. Lett.} {\bf B723} (2013) 307--313,
  [\href{http://xxx.lanl.gov/abs/1301.4965}{{\tt arXiv:1301.4965}}].

\bibitem{Djouadi:2013qya}
A.~Djouadi and G.~Moreau, {\it {The couplings of the Higgs boson and its CP
  properties from fits of the signal strengths and their ratios at the 7+8 TeV
  LHC}},  {\em Eur. Phys. J.} {\bf C73} (2013), no.~9 2512,
  [\href{http://xxx.lanl.gov/abs/1303.6591}{{\tt arXiv:1303.6591}}].

\bibitem{Degrassi:2012ry}
G.~Degrassi, S.~Di~Vita, J.~Elias-Miro, J.~R. Espinosa, G.~F. Giudice,
  G.~Isidori, and A.~Strumia, {\it {Higgs mass and vacuum stability in the
  Standard Model at NNLO}},  {\em JHEP} {\bf 08} (2012) 098,
  [\href{http://xxx.lanl.gov/abs/1205.6497}{{\tt arXiv:1205.6497}}].

\bibitem{Buttazzo:2013uya}
D.~Buttazzo, G.~Degrassi, P.~P. Giardino, G.~F. Giudice, F.~Sala, A.~Salvio,
  and A.~Strumia, {\it {Investigating the near-criticality of the Higgs
  boson}},  {\em JHEP} {\bf 12} (2013) 089,
  [\href{http://xxx.lanl.gov/abs/1307.3536}{{\tt arXiv:1307.3536}}].

\bibitem{Bertone:2004pz}
G.~Bertone, D.~Hooper, and J.~Silk, {\it {Particle dark matter: Evidence,
  candidates and constraints}},  {\em Phys. Rept.} {\bf 405} (2005) 279--390,
  [\href{http://xxx.lanl.gov/abs/hep-ph/0404175}{{\tt hep-ph/0404175}}].

\bibitem{Goudelis:2013uca}
A.~Goudelis, B.~Herrmann, and O.~Stål, {\it {Dark matter in the Inert Doublet
  Model after the discovery of a Higgs-like boson at the LHC}},  {\em JHEP}
  {\bf 09} (2013) 106, [\href{http://xxx.lanl.gov/abs/1303.3010}{{\tt
  arXiv:1303.3010}}].

\bibitem{Aranda:2012bv}
A.~Aranda, C.~Bonilla, and J.~L. Diaz-Cruz, {\it {Three generations of
Higgses
  and the cyclic groups}},  {\em Phys. Lett.} {\bf B717} (2012) 248--251,
  [\href{http://xxx.lanl.gov/abs/1204.5558}{{\tt arXiv:1204.5558}}].

\bibitem{Varzielas:2015joa}
I.~de~Medeiros~Varzielas, O.~Fischer, and V.~Maurer, {\it {$ {\mathbb{A}}_4 $
  symmetry at colliders and in the universe}},  {\em JHEP} {\bf 08} (2015)
080,
  [\href{http://xxx.lanl.gov/abs/1504.03955}{{\tt arXiv:1504.03955}}].

\bibitem{Moretti:2015tva}
S.~Moretti, D.~Rojas, and K.~Yagyu, {\it {Enhancement of the $H^\pm W^\mp Z$
  vertex in the three scalar doublet model}},  {\em JHEP} {\bf 08} (2015) 116,
  [\href{http://xxx.lanl.gov/abs/1504.06432}{{\tt arXiv:1504.06432}}].

\bibitem{Ivanov:2012ry}
I.~P. Ivanov and E.~Vdovin, {\it {Discrete symmetries in the
  three-Higgs-doublet model}},  {\em Phys. Rev.} {\bf D86} (2012) 095030,
  [\href{http://xxx.lanl.gov/abs/1206.7108}{{\tt arXiv:1206.7108}}].

\bibitem{Maniatis:2015kma}
M.~Maniatis, D.~Mehta, and C.~M. Reyes, {\it {Stability and symmetry breaking
  in a three-Higgs-doublet model with lepton family symmetry
  O(2)⊗$\mathbb{Z}_2$}},  {\em Phys. Rev.} {\bf D92} (2015), no.~3 035017,
  [\href{http://xxx.lanl.gov/abs/1503.05948}{{\tt arXiv:1503.05948}}].

\bibitem{Moretti:2015cwa}
S.~Moretti and K.~Yagyu, {\it {Constraints on Parameter Space from Perturbative
  Unitarity in Models with Three Scalar Doublets}},  {\em Phys. Rev.} {\bf D91}
  (2015) 055022, [\href{http://xxx.lanl.gov/abs/1501.06544}{{\tt
  arXiv:1501.06544}}].

\bibitem{Ivanov:2012fp}
I.~P. Ivanov and E.~Vdovin, {\it {Classification of finite reparametrization
  symmetry groups in the three-Higgs-doublet model}},  {\em Eur. Phys. J.} {\bf
  C73} (2013), no.~2 2309, [\href{http://xxx.lanl.gov/abs/1210.6553}{{\tt
  arXiv:1210.6553}}].

\bibitem{Pramanick:2015qga}
S.~Pramanick and A.~Raychaudhuri, {\it {An A4-based see-saw model for realistic
  neutrino mass and mixing}},  \href{http://xxx.lanl.gov/abs/1508.02330}{{\tt
  arXiv:1508.02330}}.

\bibitem{Keus:2015xya}
V.~Keus, S.~F. King, S.~Moretti, and D.~Sokolowska, {\it {Observable Heavy
  Higgs Dark Matter}},  {\em JHEP} {\bf 11} (2015) 003,
  [\href{http://xxx.lanl.gov/abs/1507.08433}{{\tt arXiv:1507.08433}}].

\bibitem{Keus:2014jha}
V.~Keus, S.~F. King, S.~Moretti, and D.~Sokolowska, {\it {Dark Matter with Two
  Inert Doublets plus One Higgs Doublet}},  {\em JHEP} {\bf 11} (2014) 016,
  [\href{http://xxx.lanl.gov/abs/1407.7859}{{\tt arXiv:1407.7859}}].

\bibitem{Keus:2013hya}
V.~Keus, S.~F. King, and S.~Moretti, {\it {Three-Higgs-doublet models:
  symmetries, potentials and Higgs boson masses}},  {\em JHEP} {\bf 01} (2014)
  052, [\href{http://xxx.lanl.gov/abs/1310.8253}{{\tt arXiv:1310.8253}}].

\bibitem{Bhattacharyya:2012ze}
G.~Bhattacharyya, P.~Leser, and H.~Pas, {\it {Novel signatures of the Higgs
  sector from S3 flavor symmetry}},  {\em Phys. Rev.} {\bf D86} (2012) 036009,
  [\href{http://xxx.lanl.gov/abs/1206.4202}{{\tt arXiv:1206.4202}}].

\bibitem{Bhattacharyya:2010hp}
G.~Bhattacharyya, P.~Leser, and H.~Pas, {\it {Exotic Higgs boson decay modes as
  a harbinger of $S_3$ flavor symmetry}},  {\em Phys. Rev.} {\bf D83} (2011)
  011701, [\href{http://xxx.lanl.gov/abs/1006.5597}{{\tt arXiv:1006.5597}}].

\bibitem{Barradas-Guevara:2014yoa}
E.~Barradas-Guevara, O.~Félix-Beltrán, and E.~Rodríguez-Jáuregui, {\it
  {Trilinear self-couplings in an S(3) flavored Higgs model}},  {\em Phys.
  Rev.} {\bf D90} (2014), no.~9 095001,
  [\href{http://xxx.lanl.gov/abs/1402.2244}{{\tt arXiv:1402.2244}}].

\bibitem{Kubo:2004ps}
J.~Kubo, H.~Okada, and F.~Sakamaki, {\it {Higgs potential in minimal S(3)
  invariant extension of the standard model}},  {\em Phys. Rev.} {\bf D70}
  (2004) 036007, [\href{http://xxx.lanl.gov/abs/hep-ph/0402089}{{\tt
  hep-ph/0402089}}].

\bibitem{Koide:2005ep}
Y.~Koide, {\it {Permutation symmetry S(3) and VEV structure of flavor-triplet
  Higgs scalars}},  {\em Phys. Rev.} {\bf D73} (2006) 057901,
  [\href{http://xxx.lanl.gov/abs/hep-ph/0509214}{{\tt hep-ph/0509214}}].

\bibitem{Machado:2012ed}
A.~C.~B. Machado and V.~Pleitez, {\it {Natural Flavour Conservation in a three
  Higg-doublet Model}},  \href{http://xxx.lanl.gov/abs/1205.0995}{{\tt
  arXiv:1205.0995}}.

\bibitem{Harrison:2003aw}
P.~F. Harrison and W.~G. Scott, {\it {Permutation symmetry, tri - bimaximal
  neutrino mixing and the S3 group characters}},  {\em Phys. Lett.} {\bf B557}
  (2003) 76, [\href{http://xxx.lanl.gov/abs/hep-ph/0302025}{{\tt
  hep-ph/0302025}}].

\bibitem{Kubo:2003iw}
J.~Kubo, A.~Mondragon, M.~Mondragon, and E.~Rodriguez-Jauregui, {\it {The
  Flavor symmetry}},  {\em Prog. Theor. Phys.} {\bf 109} (2003) 795--807,
  [\href{http://xxx.lanl.gov/abs/hep-ph/0302196}{{\tt hep-ph/0302196}}].
  [Erratum: Prog. Theor. Phys.114,287(2005)].

\bibitem{Teshima:2005bk}
T.~Teshima, {\it {Flavor mass and mixing and S(3) symmetry: An S(3) invariant
  model reasonable to all}},  {\em Phys. Rev.} {\bf D73} (2006) 045019,
  [\href{http://xxx.lanl.gov/abs/hep-ph/0509094}{{\tt hep-ph/0509094}}].

\bibitem{Koide:2006vs}
Y.~Koide, {\it {S(3) symmetry and neutrino masses and mixings}},  {\em Eur.
  Phys. J.} {\bf C50} (2007) 809--816,
  [\href{http://xxx.lanl.gov/abs/hep-ph/0612058}{{\tt hep-ph/0612058}}].

\bibitem{Chen:2004rr}
S.-L. Chen, M.~Frigerio, and E.~Ma, {\it {Large neutrino mixing and normal mass
  hierarchy: A Discrete understanding}},  {\em Phys. Rev.} {\bf D70} (2004)
  073008, [\href{http://xxx.lanl.gov/abs/hep-ph/0404084}{{\tt
  hep-ph/0404084}}]. [Erratum: Phys. Rev.D70,079905(2004)].

\bibitem{Mondragon:2007af}
A.~Mondragon, M.~Mondragon, and E.~Peinado, {\it {Lepton masses, mixings and
  FCNC in a minimal S(3)-invariant extension of the Standard Model}},  {\em
  Phys. Rev.} {\bf D76} (2007) 076003,
  [\href{http://xxx.lanl.gov/abs/0706.0354}{{\tt arXiv:0706.0354}}].

\bibitem{Fortes:2014dca}
E.~C. F.~S. Fortes, A.~C.~B. Machado, J.~Montaño, and V.~Pleitez, {\it {Scalar
  dark matter candidates in a two inert Higgs doublet model}},  {\em J. Phys.}
  {\bf G42} (2015), no.~10 105003,
  [\href{http://xxx.lanl.gov/abs/1407.4749}{{\tt arXiv:1407.4749}}].

\bibitem{Aranda:2013kq}
A.~Aranda, C.~Bonilla, F.~de~Anda, A.~Delgado, and J.~Hernandez-Sanchez, {\it
  {Higgs decay into two photons from a 3HDM with flavor symmetry}},  {\em Phys.
  Lett.} {\bf B725} (2013) 97--100,
  [\href{http://xxx.lanl.gov/abs/1302.1060}{{\tt arXiv:1302.1060}}].

\bibitem{Ivanov:2014doa}
I.~P. Ivanov and C.~C. Nishi, {\it {Symmetry breaking patterns in 3HDM}},  {\em
  JHEP} {\bf 01} (2015) 021, [\href{http://xxx.lanl.gov/abs/1410.6139}{{\tt
  arXiv:1410.6139}}].

\bibitem{Serebrov:2013tba}
A.~P. Serebrov et~al., {\it {New measurements of the neutron electric dipole
  moment}},  {\em JETP Lett.} {\bf 99} (2014) 4--8,
  [\href{http://xxx.lanl.gov/abs/1310.5588}{{\tt arXiv:1310.5588}}].

\bibitem{Chakrabarty:2014aya}
N.~Chakrabarty, U.~K. Dey, and B.~Mukhopadhyaya, {\it {High-scale validity of a
  two-Higgs doublet scenario: a study including LHC data}},  {\em JHEP} {\bf
  12} (2014) 166, [\href{http://xxx.lanl.gov/abs/1407.2145}{{\tt
  arXiv:1407.2145}}].

\bibitem{1742-6596-171-1-012028}
O.~F. Beltrán, M.~Mondragón, and E.~Rodríguez-Jáuregui, {\it Conditions for
  vacuum stability in an s 3 extension of the standard model},  {\em Journal of
  Physics: Conference Series} {\bf 171} (2009), no.~1 012028.

\bibitem{Das:2014fea}
D.~Das and U.~K. Dey, {\it {Analysis of an extended scalar sector with $S_3$
  symmetry}},  {\em Phys. Rev.} {\bf D89} (2014), no.~9 095025,
  [\href{http://xxx.lanl.gov/abs/1404.2491}{{\tt arXiv:1404.2491}}]. [Erratum:
  Phys. Rev.D91,no.3,039905(2015)].

\bibitem{Searches:2001ac}
{\bf OPAL, DELPHI, L3, ALEPH, LEP Higgs Working Group for Higgs boson searches}
  Collaboration, {\it {Search for charged Higgs bosons: Preliminary combined
  results using LEP data collected at energies up to 209-GeV}},  in {\em
  {Lepton and photon interactions at high energies. Proceedings, 20th
  International Symposium, LP 2001, Rome, Italy, July 23-28, 2001}}, 2001.
\newblock \href{http://xxx.lanl.gov/abs/hep-ex/0107031}{{\tt hep-ex/0107031}}.

\bibitem{Arhrib:2013ela}
A.~Arhrib, Y.-L.~S. Tsai, Q.~Yuan, and T.-C. Yuan, {\it {An Updated Analysis of
  Inert Higgs Doublet Model in light of the Recent Results from LUX, PLANCK,
  AMS-02 and LHC}},  {\em JCAP} {\bf 1406} (2014) 030,
  [\href{http://xxx.lanl.gov/abs/1310.0358}{{\tt arXiv:1310.0358}}].

\bibitem{Das:2015sca}
D.~Das, U.~K. Dey, and P.~B. Pal, {\it {$S_3$ symmetry and the CKM matrix}},
  \href{http://xxx.lanl.gov/abs/1507.06509}{{\tt arXiv:1507.06509}}.

\bibitem{Canales:2013cga}
F.~González~Canales, A.~Mondragón, M.~Mondragón, U.~J. Saldaña~Salazar, and
  L.~Velasco-Sevilla, {\it {Quark sector of S3 models: classification and
  comparison with experimental data}},  {\em Phys. Rev.} {\bf D88} (2013)
  096004, [\href{http://xxx.lanl.gov/abs/1304.6644}{{\tt arXiv:1304.6644}}].

\bibitem{Ma:2013zca}
E.~Ma and B.~Melic, {\it {Updated $S_{3}$ model of quarks}},  {\em Phys. Lett.}
  {\bf B725} (2013) 402--406, [\href{http://xxx.lanl.gov/abs/1303.6928}{{\tt
  arXiv:1303.6928}}].

\bibitem{Teshima:2011wg}
T.~Teshima and Y.~Okumura, {\it {Quark/lepton mass and mixing in $S_3$
  invariant model and CP-violation of neutrino}},  {\em Phys. Rev.} {\bf D84}
  (2011) 016003, [\href{http://xxx.lanl.gov/abs/1103.6127}{{\tt
  arXiv:1103.6127}}].

\bibitem{PhysRevD.16.1519}
B.~W. Lee, C.~Quigg, and H.~B. Thacker, {\it Weak interactions at very high
  energies: The role of the higgs-boson mass},  {\em Phys. Rev. D} {\bf 16}
  (Sep, 1977) 1519--1531.

\bibitem{Akeroyd:2000wc}
A.~G. Akeroyd, A.~Arhrib, and E.-M. Naimi, {\it {Note on tree level unitarity
  in the general two Higgs doublet model}},  {\em Phys. Lett.} {\bf B490}
  (2000) 119--124, [\href{http://xxx.lanl.gov/abs/hep-ph/0006035}{{\tt
  hep-ph/0006035}}].

\bibitem{Horejsi:2005da}
J.~Horejsi and M.~Kladiva, {\it {Tree-unitarity bounds for THDM Higgs masses
  revisited}},  {\em Eur. Phys. J.} {\bf C46} (2006) 81--91,
  [\href{http://xxx.lanl.gov/abs/hep-ph/0510154}{{\tt hep-ph/0510154}}].

\bibitem{Gorczyca:2011he}
B.~Gorczyca and M.~Krawczyk, {\it {Tree-Level Unitarity Constraints for the
  SM-like 2HDM}},  \href{http://xxx.lanl.gov/abs/1112.5086}{{\tt
  arXiv:1112.5086}}.

\bibitem{Branchina:2013jra}
V.~Branchina and E.~Messina, {\it {Stability, Higgs Boson Mass and New
  Physics}},  {\em Phys. Rev. Lett.} {\bf 111} (2013) 241801,
  [\href{http://xxx.lanl.gov/abs/1307.5193}{{\tt arXiv:1307.5193}}].

\bibitem{PhysRevD.91.013003}
V.~Branchina, E.~Messina, and M.~Sher, {\it Lifetime of the electroweak
vacuum
  and sensitivity to planck scale physics},  {\em Phys. Rev. D} {\bf 91}
(Jan,
  2015) 013003.

\bibitem{Grimus:2008nb}
W.~Grimus, L.~Lavoura, O.~M. Ogreid, and P.~Osland, {\it {The Oblique
  parameters in multi-Higgs-doublet models}},  {\em Nucl. Phys.} {\bf B801}
  (2008) 81--96, [\href{http://xxx.lanl.gov/abs/0802.4353}{{\tt
  arXiv:0802.4353}}].

\bibitem{Baak:2013ppa}
M.~Baak and R.~Kogler, {\it {The global electroweak Standard Model fit after
  the Higgs discovery}},  in {\em {Proceedings, 48th Rencontres de Moriond on
  Electroweak Interactions and Unified Theories}}, pp.~349--358, 2013.
\newblock \href{http://xxx.lanl.gov/abs/1306.0571}{{\tt arXiv:1306.0571}}.
\newblock [,45(2013)].

\bibitem{Gunion:2002zf}
J.~F. Gunion and H.~E. Haber, {\it {The CP conserving two Higgs doublet model:
  The Approach to the decoupling limit}},  {\em Phys. Rev.} {\bf D67} (2003)
  075019, [\href{http://xxx.lanl.gov/abs/hep-ph/0207010}{{\tt
  hep-ph/0207010}}].

\bibitem{Aad:2014eha}
{\bf ATLAS} Collaboration, G.~Aad et~al., {\it {Measurement of Higgs boson
  production in the diphoton decay channel in pp collisions at center-of-mass
  energies of 7 and 8 TeV with the ATLAS detector}},  {\em Phys. Rev.} {\bf
  D90} (2014), no.~11 112015, [\href{http://xxx.lanl.gov/abs/1408.7084}{{\tt
  arXiv:1408.7084}}].

\bibitem{Khachatryan:2014jba}
{\bf CMS} Collaboration, V.~Khachatryan et~al., {\it {Precise
determination of
  the mass of the Higgs boson and tests of compatibility of its couplings
with
  the standard model predictions using proton collisions at 7 and 8 $\,\text
  {TeV}$}},  {\em Eur. Phys. J.} {\bf C75} (2015), no.~5 212,
  [\href{http://xxx.lanl.gov/abs/1412.8662}{{\tt arXiv:1412.8662}}].

\bibitem{Ade:2013zuv}
{\bf Planck} Collaboration, P.~A.~R. Ade et~al., {\it {Planck 2013 results.
  XVI. Cosmological parameters}},  {\em Astron. Astrophys.} {\bf 571} (2014)
  A16, [\href{http://xxx.lanl.gov/abs/1303.5076}{{\tt arXiv:1303.5076}}].

\bibitem{Belanger:2013oya}
G.~Belanger, F.~Boudjema, A.~Pukhov, and A.~Semenov, {\it {micrOMEGAs$_3$: A
  program for calculating dark matter observables}},  {\em Comput. Phys.
  Commun.} {\bf 185} (2014) 960--985,
  [\href{http://xxx.lanl.gov/abs/1305.0237}{{\tt arXiv:1305.0237}}].

\bibitem{Aprile:2012nq}
{\bf XENON100} Collaboration, E.~Aprile et~al., {\it {Dark Matter Results from
  225 Live Days of XENON100 Data}},  {\em Phys. Rev. Lett.} {\bf 109} (2012)
  181301, [\href{http://xxx.lanl.gov/abs/1207.5988}{{\tt arXiv:1207.5988}}].

\bibitem{Akerib:2013tjd}
{\bf LUX} Collaboration, D.~S. Akerib et~al., {\it {First results from the LUX
  dark matter experiment at the Sanford Underground Research Facility}},  {\em
  Phys. Rev. Lett.} {\bf 112} (2014) 091303,
  [\href{http://xxx.lanl.gov/abs/1310.8214}{{\tt arXiv:1310.8214}}].

\bibitem{Ferreira:2009jb}
P.~M. Ferreira and D.~R.~T. Jones, {\it {Bounds on scalar masses in two Higgs
  doublet models}},  {\em JHEP} {\bf 08} (2009) 069,
  [\href{http://xxx.lanl.gov/abs/0903.2856}{{\tt arXiv:0903.2856}}].

\bibitem{Sher:1988mj}
M.~Sher, {\it {Electroweak Higgs Potentials and Vacuum Stability}},  {\em Phys.
  Rept.} {\bf 179} (1989) 273--418.

\bibitem{PAKVASA197861}
S.~Pakvasa and H.~Sugawara, {\it Discrete symmetry and cabibbo angle},  {\em
  Physics Letters B} {\bf 73} (1978), no.~1 61 -- 64.

\bibitem{Ahriche:2015mea}
A.~Ahriche, G.~Faisel, S.-Y. Ho, S.~Nasri, and J.~Tandean, {\it {Effects
of two
  inert scalar doublets on Higgs boson interactions and the electroweak phase
  transition}},  {\em Phys. Rev.} {\bf D92} (2015), no.~3 035020,
  [\href{http://xxx.lanl.gov/abs/1501.06605}{{\tt arXiv:1501.06605}}].

\bibitem{Branco:2011iw}
G.~C. Branco, P.~M. Ferreira, L.~Lavoura, M.~N. Rebelo, M.~Sher, and J.~P.
  Silva, {\it {Theory and phenomenology of two-Higgs-doublet models}},  {\em
  Phys. Rept.} {\bf 516} (2012) 1--102,
  [\href{http://xxx.lanl.gov/abs/1106.0034}{{\tt arXiv:1106.0034}}].

\bibitem{Djouadi:2005gi}
A.~Djouadi, {\it {The Anatomy of electro-weak symmetry breaking. I: The Higgs
  boson in the standard model}},  {\em Phys. Rept.} {\bf 457} (2008) 1--216,
  [\href{http://xxx.lanl.gov/abs/hep-ph/0503172}{{\tt hep-ph/0503172}}].






























\end{thebibliography}\endgroup
\end{document}